\documentclass[pra,twocolumn,aps,floatfix,10pt,superscriptaddress,a4paper,longbibliography]{revtex4-2}

\usepackage{amsmath, amsfonts, amsthm, amssymb} 
\usepackage{dsfont}
\usepackage{physics}
\usepackage[hidelinks]{hyperref}

\usepackage{listings} 

\usepackage[english]{babel} 

\usepackage[dvipsnames]{xcolor}
\usepackage{graphicx} 
\graphicspath{{figures/}{./}} 

\usepackage{enumitem} 
\setlist{noitemsep} 

\usepackage{empheq}
\usepackage{fancybox}

\renewcommand{\vec}{\mathbf}

\newcommand{\be}{\begin{eqnarray}}
\newcommand{\ee}{\end{eqnarray}}
\newcommand{\bse}{\begin{subequations}}
	\newcommand{\ese}{\end{subequations}}

\newcommand{\bnum}{\begin{enumerate}}
	\newcommand{\enum}{\end{enumerate}}

\newcommand{\bit}{\begin{itemize}}
	\newcommand{\eit}{\end{itemize}}

\newcommand{\bc}{\begin{cases}}
	\newcommand{\ec}{\end{cases}}

\newcommand{\bpm}{\begin{pmatrix}}
	\newcommand{\epm}{\end{pmatrix}}

\newcommand{\bvm}{\begin{vmatrix}}
	\newcommand{\evm}{\end{vmatrix}}

\newcommand{\mrm}{\mathrm}

\newcommand{\eps}{\epsilon}

\newcommand{\gl}{\lambda}
\newcommand{\gk}{\kappa}
\newcommand{\go}{\omega}
\newcommand{\gt}{\theta}

\newcommand{\Gd}{\Delta}

\newcommand{\p}{\partial}
\newcommand{\f}{\frac}
\newcommand{\diff}{\mrm{d}}
\newcommand{\iy}{\infty}

\newcommand{\R}{\mathbb{R}}

\newcommand{\Z}{\mathbb{Z}}

\newcommand{\lan}{\langle}
\newcommand{\ran}{\rangle}

\usepackage[utf8]{inputenc} 
\usepackage[T1]{fontenc} 
\usepackage[eulergreek]{sansmath}

\usepackage{pgf,tikz}
\usepackage{pgfplots}
\usepgfplotslibrary{groupplots}

\pgfplotsset{
    compat=1.18,
    every axis/.append style={font=\small, axis line style={thick}},
	every axis label/.append style={font=\normalsize}, 
	tick align=outside, 
	tick pos=left,
	/tikz/font={\sffamily}, 
	xlabel near ticks, 
	ylabel near ticks, 
	every axis/.append style={line width=1pt}, 
	every tick/.append style={line width=1pt,color=black}, 
	every tick label/.append style={font={\sansmath\sffamily\small}},
}

\usepackage{bbm}

\newcommand{\vx}{\vec{x}}

\newcommand{\vk}{\vec{k}}

\newcommand{\vp}{\vec{p}}
\newcommand{\vq}{\vec{q}}

\newcommand{\crit}{\text{p}}
\newcommand{\kcrit}{k_\text{p}}
\newcommand{\ocrit}{\go_\text{p}}
\newcommand{\vkcrit}{\vec{k}_\text{p}}

\newcommand{\nlaplace}{\tilde{n}}

\newcommand{\cov}{\mathcal{C}}

\begin{document}

\title{Disorder-induced broadening of quantum momentum distribution}

\author{Vili Heinonen}
\affiliation{
	Department of Mathematics and Statistics, University of Helsinki, P.O. Box 68, FI-00014 Helsingin yliopisto, Finland
}
\email{vili.heinonen@helsinki.fi}

\author{Jani Lukkarinen}
\affiliation{
	Department of Mathematics and Statistics, University of Helsinki, P.O. Box 68, FI-00014 Helsingin yliopisto, Finland
}
\email{jani.lukkarinen@helsinki.fi}

\begin{abstract}
	We study the long-time behavior of a non-interacting two-dimensional quantum gas in a weak random potential with long-range correlations. Any peaked initial momentum distribution will eventually become isotropic and broaden due to scattering events with the random potential. We derive an expression for the long-time average of the momentum distribution and test it against computer simulations. We also discuss momentum isotropization and spatial diffusion. 
\end{abstract}

\maketitle 
\thispagestyle{empty}


\section{Introduction}\label{sec:introduction}
Whether studying the transport of charged particles \cite{Baranovski2006} or light \cite{Vynck2023}, the presence of disorder plays a fundamental role in the physics of the system. Some of the hallmark disorder-induced phenomena include, e.g., Anderson \cite{Anderson1958,Abrahams1979} and weak localization \cite{Gorkov1996,Akkermans1985} manifesting as inhibition (Anderson) or reduction (weak) of diffusion due to interference of multiply scattered waves. Similar mechanisms underlie the appearance of backward \cite{Jendrzejewski2012} and forward scattering peaks \cite{Karpiuk2012} in the momentum spectrum observed as sharply increased wave intensity in the opposite (backward) or same (forward) direction with the incident wave. Studying quantum transport in disordered environments has become experimentally accessible with the advent of cold atom physics revolutionizing experimental study of quantum transport. 

One of the major problems in previously studied solid-state systems is that the particles are interacting \cite{Baranovski2006} leading to significantly more complicated physics. Conversely, in cold atom clouds these interactions can be readily suppressed e.g. by using spin-polarized fermions \cite{Kondov2011three} or by tuning the Feshback resonances via external magnetic fields \cite{Chin2010feshbach}. In these systems the spatially disordered background can be realized via interactions with light obtained by reflecting a laser source with a rough plate \cite{Goodman2007speckle,Kuhn2007}. Using additional confining optical potentials also allows studying systems in effective lower dimensions \cite{Petrov2004low}. Furthermore, time series of momentum data can be accurately recorded in experiments by using e.g. momentum space through time-of-flight absorption imaging \cite{Labeyrie2012}.

In the non-interacting regime the momentum of a collection of particles is conserved in the absence of an external potential. Introducing a weak random potential at the energy scale $\eps$ will induce momentum isotropization through scattering events \cite{Jendrzejewski2012,Labeyrie2012,Plisson2013}.
The linear Boltzmann equation governing this isotropization has been previously derived in the kinetic scaling limit $\eps \to 0$ \cite{Spohn1977,Erdos2000}. 
At this limit these scattering events are elastic preserving the energy of the incident wave. However, keeping $\eps$ finite will induce a broadening of the momentum (kinetic energy) spectrum of the particle. In this Article we derive an expression for the momentum (kinetic energy) distribution assuming that the correlations in the random environment are long-range compared to the wavelength of the incident wave. 

The Article is organized as follows: In Sec.~\ref{sec:model} we introduce the underlying dynamics and describe the random environment, while Sec.~\ref{sec:averaged_dynamics} discusses ensemble averaged dynamics. The main results are presented in Sec.~\ref{sec:long-time_properties}, where we derive the expression for momentum broadening as well as discuss isotropization. In Secs.~\ref{sec:boltzmann} and \ref{sec:diffusion} we calculate the transport coefficients using methods described in \cite{Erdos2008}. Sec.~\ref{sec:conclusion} concludes the Article and discusses some ramifications for thermalization in quantum mechanical systems.

\section{The model}\label{sec:model}
Consider the Schrödinger equation for non-interacting particles with mass $m$
in a two-dimensional periodic domain of size $ L \times L $. The system is initialized with a single wave packet $ \psi (\vx)  \propto \exp(i \vkcrit \cdot \vx) $ with $|\vkcrit| =: \kcrit $ with $1/\kcrit \ll L$. We choose the units s.t. the length scale is given by $ x_0 = 1/k_0 $, mass by $ m_0 = m $ and time by $ t_0^{-1} = \hbar k_0^2 /m $, where $k_0$ is a length scale comparable to $\kcrit$.
The dynamics for $ \psi(t,\vx) $ in $\vx$-space are given by
\begin{equation}\label{eq:schrodinger_dynamics}
	i \partial_t \psi(t,\vx) =  -\f{1}{2} \nabla^2 \psi(t,\vx) +  \eps V(\vx) \psi (t,\vx) ,
\end{equation}
where we take $ V(\vx) $ to be an isotropic homogeneous random field
\begin{equation}\label{eq:potential_energy}
	V(\vx) = \sum\nolimits_{\vk} V_\vk e^{i \vk \cdot x}
\end{equation}
with zero mean ($\lan V(\vx) \ran = 0$). Throughout this Article, the angle brackets $\lan \cdot \ran$ denote the average over \emph{realizations} of the random field $V$. 

The potential strength is set by the parameter $\eps $, which is assumed to be small compared to the kinetic energy of the initial wave packet. 
We define the 2-point covariance function
\begin{equation}
	\cov (\vx) = \lan V(\vx) V(0) \ran = \sum\nolimits_\vk \cov_\vk e^{i \vk \cdot \vx},
\end{equation}
where
\begin{equation}\label{eq:potential_independence}
	\cov_\vk = \left \lan | V_\vk|^2 \right  \ran
\end{equation}
due to the homogeneity of $V$ (See App.~\ref{sec:random_fields}). 
We assume that $\cov$ describes long-range correlations peaking at $\vk = 0$ and having a natural length scale $\zeta$ for which $\zeta \kcrit \ll 1$. This corresponds to the assumption that the random potential $V$ varies at large length scales compared to the wavelength of the initial wave packet. Furthermore, we assume that the potential is peaked at $\vk = 0$ so that it is effectively described by a Gaussian
\begin{equation}\label{eq:covariance-approximation}
	\cov_\vk \approx C_\cov e^{-\f 1 2 \zeta^2 k^2},
\end{equation}
where the normalization constant $C_\cov$ is set by
\begin{equation}
	\langle V(\vx )^2 \rangle = \langle V(0)^2 \rangle = \sum\nolimits_{\vk} \cov_\vk = 1.
\end{equation}

In $ \vk $-space the dynamics Eq.~\eqref{eq:schrodinger_dynamics} can be written as 
\begin{equation}\label{eq:1schrodinger_kspace}
	\partial_t  \psi _\vk
	= -\f i 2 k^2  \psi_\vk -i \eps \sum\nolimits_{\vq}  V_{\vk - \vq}  \psi_{\vq}.
\end{equation}
We write the dynamics in the interaction picture by defining $ \phi_{\vec{k}} = {\psi}_\vk \exp(i k^2 t/2) $. The time evolution for $\phi_{\vk}$ becomes 
%
\begin{empheq}[box=\ovalbox]{equation}\label{eq:schrodinger_kspace}
	\partial_t  \phi _\vk = \eps \sum\nolimits_{\vq}K_{kq}  V_{\vk - \vq}  \phi_{\vq},
\end{empheq}
where 
\begin{subequations}
	\begin{align}
		K_{kq}(t) &= -i \exp(i t T_{kq}), \\
		T_{kq} &= \frac{k^2 -q^2}{2}.
	\end{align}
\end{subequations}
The interaction kernel $K$ has the symmetries  
\begin{equation}
	\label{eq:k-symmetries}
	K_{kq}(t)^* = - K_{kq}(-t) = - K_{qk}(t)
\end{equation}
and the multiplication property
\begin{equation}
	\label{eq:k-addition}
	K_{kq}(t_1) K_{kq}(t_2)^* = i K_{kq}(t_1 - t_2). 
\end{equation} 
Note that $e^{-i k^2 t/2}$ is the $k$ space representation $\bra{k} U_\text{f}(t) \ket{k}$ of the time evolution operator $U_\text{f}$ of the free particle. 

We initialize the fields $\phi$ with a random global phase factor $e^{i S}$, where $S$ is uniformly distributed and independent of the potential $V$. The addition of this phase factor leaves the dynamics invariant and
implies that terms of the form 
\begin{equation}\label{eq:potential_field_correlation}
	\left \lan  V_{\vk_1} \ldots V_{\vk_n}  \phi_{\vq_1} \ldots \phi_{\vq_m} \right \ran  = 0,
\end{equation}
if the list $(\phi_{\vq_1},\ldots, \phi_{\vq_m})$ does not contain an equal number of fields and their complex conjugates. In particular, we have $\lan \phi_\vk \ran = 0$. 

\begin{figure}
	\centering
	\includegraphics[width=81.5mm]{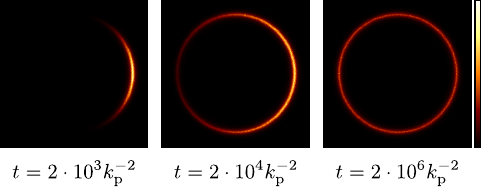}
	\caption{Time evolution of disorder-averaged density field $n = \lan |\phi|^2\ran$ generated by Eq.~\eqref{eq:schrodinger_kspace} showing momentum isotropization. In the first panel the brightest spot corresponds to the initial peak. Dynamics are averaged over 50 realizations and the parameters are $\eps = \frac{1}{32} \kcrit^2$, $\zeta=12 \kcrit^{-1}$. The densities are normalized s.t. the different panels are on the same scale. 
		See App.~\ref{sec:numerical-methods} for details on the simulations.
	}
	\label{fig:evolution}
\end{figure}

Dynamics of Eq.~\eqref{eq:schrodinger_kspace} leads to momentum isotropization as $t \to \iy$. Fig.~\ref{fig:evolution} shows snapshots from simulations illustrating this process. 

\section{Averaged dynamics}\label{sec:averaged_dynamics}

In this Section we will look at the time evolution of the average particle number at $\vk$
\begin{equation}\label{eq:wigner_functions}
	n_\vk := \left \langle \phi^*_{\vk} \phi_\vk \right \rangle,
\end{equation}
which is given by 
\begin{equation}\label{eq:wigner_dynamics}
	\partial_t n_\vk = \langle  \phi^*_\vk \partial_t \phi_{\vec{k}} \rangle  + \text{c.c.} =   \sum\nolimits_{\vq} 2 \eps\Re \left[ K_{kq} \left \langle  V_{\vk - \vq}  \phi_{\vq} \phi^*_\vk \right \rangle   \right],
\end{equation}
where $ \text{c.c.} $ stands for \emph{complex conjugate}.
%
%
The time evolution of the triad on the RHS can be written as
\begin{equation}\label{eq:triad_time_evolution}
	\begin{split}
		\partial_t \left \langle  V_{\vk - \vq}  \phi_{\vq} \phi^*_\vk \right \rangle
		&= \eps \sum\nolimits_{\vp} 
		\left [  K_{qp}
		\left \lan  V _{\vq - \vp } \phi_\vp  V _{\vk - \vq} \phi^* _\vk \right \ran \right. \\ &\left.
		+  K_{kp}^* \left \lan  V^* _{\vk - \vp } \phi^*_\vp  V _{\vk - \vq} \phi _\vq \right \ran 
		\right].
	\end{split}
\end{equation}
We expand the moments to cumulants on the RHS generating 4th and 2nd order cumulants (See App.~\ref{sec:cumulants} and Eq.~\eqref{eq:moments2cumulants} therein). There are no 3rd order cumulants in this expansion because the trailing 1st order cumulant (moment) is zero both for the coefficients of the field $\phi$ and the potential $V$. This is a consequence of the homogeneity and zero average of $V$ and the added phase degrees of freedom for $\phi$ discussed earlier. 
For the same reason 2nd order cumulants can be written as second order moments. 

Assuming that the effect of the 4th order cumulants is negligible (small $\eps$ expansion) gives a sum of products of 2nd order cumulants. 
Eq.~\eqref{eq:potential_field_correlation} implies that the 2nd order cumulants between the field and the potential are zero
i.e. the only terms that contribute are terms with two coefficients of the background potential and two coefficients of the field $ \phi $. Now
\begin{equation}\label{eq:1triad_evolution}
	\begin{split}
		\partial_t \left \langle  V_{\vk - \vq}  \phi_{\vq} \phi^*_\vk \right \rangle
		&= \eps \sum\nolimits_{\vp} 
		\left [ K_{qp}
		 \left \lan  V _{\vq - \vp } V _{\vk - \vq}   \right \ran  \left \lan \phi_\vp \phi^* _\vk \right \ran
		\right. \\ &\left.
		+K_{kp}^* \left \lan  V _{\vp - \vk }   V _{\vk - \vq}\right \ran \left \lan \phi^*_\vp \phi _\vq \right \ran 
		\right ]
	\end{split}
\end{equation}
Here we have used the fact that $ V $ is a real field i.e. $ {V}^*_\vk = {V}_{-\vk} $. Using the homogeneity of the field $V$ gives
\begin{equation}\label{eq:triad_evolution}
		\partial_t \left \langle  V_{\vk - \vq}  \phi_{\vq} \phi^*_\vk \right \rangle
		=  \eps 
		\left (
		n_\vq 
		- n_\vk 
		\right )\cov _{\vk - \vq} K_{kq}^* ,
\end{equation}
where we have used $K_{qk} = -K_{kq}^*$. For more details, see App.~\ref{app:cumulant_expansion}.


Integrating Eq.~\eqref{eq:triad_evolution} and substituting into Eq.~\eqref{eq:wigner_dynamics} gives 
\begin{equation}\label{eq:1wigner_dynamics}
	\begin{split}
		&\partial_t n_\vk(t) = 2 \eps \sum\nolimits_{\vq} \Re\left[ \left \langle  V_{\vk - \vq}  \phi_{\vq} \phi^*_\vk \right \rangle_{t=0} K_{kq}(0) \right] \\
		&+2\eps^2 \Re \sum\nolimits_{\vq} \cov _{\vk-\vq} \int_{0}^{t} \dd{t'} i K_{kq}(t-t')\Gd n_{\vq, \vk} (t'),
	\end{split}
\end{equation}
where we used the addition property Eq.~\eqref{eq:k-addition} of $K$ and $\Gd n_{\vq,\vk} = n_\vq - n_\vk$.
We notice that at time $t=0$ the coefficients $\phi_\vk$ are deterministic and $\lan  V_\vk \ran = 0$ implying that the first term is zero. Therefore, the truncated hierarchy leads to the the evolution equation
\begin{empheq}[box=\ovalbox]{equation}
	\label{eq:density-dynamics}
	\partial_t n_\vk(t) = \sum\nolimits_{\vq} 2 \eps^2 \cov_{\vk-\vq} \int_{0}^{t} \diff t' \cos(T_{kq} (t -t')) \Gd n_{\vq, \vk} (t').
\end{empheq}

\section{Long-time properties}\label{sec:long-time_properties}
In this section, we will study how the density $n$ spreads on a ring in momentum-space around the initial momentum $\kcrit$. To this end, we will consider the continuum limit ($L \to \infty$) of the previous periodic system. The corresponding continuum fields $n(\vk)$ and $\cov(\vk)$ are subject to normalization 
\begin{subequations}
	\begin{align}
		\int \dd{\vk} n(t,\vk) &= 1, \\
		\int \dd{\vk} \cov(\vk) &= 1.
	\end{align}
\end{subequations}
For the approximate covariance function this implies that 
\begin{equation}\label{eq:covariance-continuous}
	\cov(\vk) = \frac{\zeta^2}{2\pi} e^{-\f 1 2 \zeta^2 k^2}.
\end{equation}
Explicitly, the above continuum limits are obtained from the corresponding lattice quantities via $n(t,\vk) = \lim_{L \to \infty}  n_\vk(t)  / \Delta k^2$ and
$\cov (\vk) = \lim_{L \to \infty}  \cov_\vk / \Delta k^2$, where $\Delta k$ is the discrete volume element $2\pi/L$. Per previous discussion, we assume an initial condition $n(0,\vk) = \delta(\vk - \vkcrit)$, where $\delta$ is the Dirac delta. 

In the continuum description, Eq.~\eqref{eq:density-dynamics} reads
\begin{equation}
	\label{eq:dynamics-continuum}
	\begin{split}
		\p_t n(t,\vk) &= \int \dd{\vq} 2\eps^2 \cov (\vk - \vq) 
		\\ & \times \int_{0}^{t} \dd{t'} \cos(T(k,q) (t-t')) \Gd n(\vq,\vk,t').
	\end{split}
\end{equation}

Next, we take the Laplace transform
\begin{equation}
	\mathcal{L}[n(t,\vk)](s) = \int_0^{\infty} \diff t e^{-st} n(t,\vk)
\end{equation}
of Eq.~\eqref{eq:dynamics-continuum}. 
On the LHS we use the property
\begin{equation}
	\mathcal L [\p_t n(t,\vk)](s) = \bar n(s,\vk) - n(t=0,\vk),
\end{equation}
where we define
\begin{equation}
	\bar n(s, \vk) := s \mathcal{L}[n(t,\vk)](s),
\end{equation}
whereas on the RHS of Eq.~\eqref{eq:dynamics-continuum} we use the convolution property
\begin{equation}
	\mathcal{L}[f*g](s) = \mathcal{L}[f](s) \mathcal{L}[g](s),
\end{equation}
where 
\[ (f*g)(t) = \int_{0}^{t}\dd{t'} f(t-t') g(t') \]
is the convolution.
Since 
\[ \mathcal{L} \left [\cos(t T(k,q))\right ](s)  = \frac{s}{T(k,q)^2 + s^2}\,, \]
we finally have
\begin{empheq}[box=\ovalbox]{equation}
	\label{eq:laplace_transform}
	\bar  n(s,\vk) - n(0,\vk) =  \int \dd{\vq} \frac{2 \eps^2 \cov (\vk-\vq) }{T(k,q)^2 + s^2} \Gd \bar n(s,\vq,\vk),
\end{empheq}
where, again, $\Gd \bar n(s,\vq,\vk) = \bar n (s,\vq) - \bar n (s,\vk)$.

The definition of $\bar n$ is motivated by the property
\begin{equation}
	\lim_{s \to 0^+} \bar n(s,\vk) = \lim_{T \to \infty} \frac{1}{T} \int_{0}^{T} \dd{t} n(t,\vk),
\end{equation}
which holds in a distributional sense if $\int \dd{\vk} f(\vk) n(t,\vk)$ is bounded for suitably nice test functions $f$. 
Ultimately, this identity will be used to extract time averaged behavior of $n$.
Multiplying Eq.~\eqref{eq:laplace_transform} by $s$ and sending $s \to 0$ gives 0 on the LHS and on the RHS we use the identity
\begin{equation}
	\lim_{s \to 0} \frac{s}{s^2 + x^2} = \pi \delta(x).
\end{equation}
This gives the equation
\begin{equation}
	\label{eq:second-limit}
	\int \dd{\vq} \delta(T(k,q)) \cov(\vk-\vq) \left ( \bar n (0,\vq) - \bar n (0,\vk) \right ) = 0,
\end{equation}
which, since $T(k,q) = \frac{k^2}{2} - \frac{q^2}{2} $ and $\cov$ is positive and isotropic, is exactly the isotropy condition for $\bar n(0,\vk)$. This implies that the long-time dynamics Eq.~\eqref{eq:dynamics-continuum} will eventually isotropize the field $n$. 

\subsection{Momentum broadening}\label{sec:momentum-broadening}
In order to capture the radial behavior of $\lim_{s\to 0} \bar n$,
we integrate Eq.~\eqref{eq:laplace_transform} over the circle fixed by the radius $k$. We introduce radial variables 
\begin{equation}
	\bar n(s,k) = \int \dd{\vk'} \frac{\delta(k-k')}{k} n(s,\vk')
\end{equation}
and
\begin{equation}\label{eq:polar-covariance}
	\begin{split}
		\cov (k,q) &= \int \dd{\vk'} \frac{\delta(k-k')}{k} \cov(\vk' - \vq) \\
		&= \int \dd{\vq'} \frac{\delta(q-q')}{q} \cov(\vk - \vq').
	\end{split}
\end{equation}
The normalization conditions read $\int_{0}^{\iy} \dd{k} k \bar n (s,k) = 1$ and $\int_{0}^{\iy} \dd{k} k \cov(k,q) = \int_{0}^{\iy} \dd{q} q \cov (k,q) = 1$. 
This form of $\cov$ is possible because $\cov$ is isotropic and is therefore a function of the squared modulus $|\vk -\vq|^2 = k^2 + q^2 - 2 k q \cos \gt$, where $\gt$ is the angle between $\vk$ and $\vq$. This implies also the symmetry $\cov (k,q) = \cov (q,k)$.
Upon integration, the LHS gives 
\begin{equation}
	\int \dd{\vk'} \frac{\delta(k-k')}{k} (\bar n (s,\vk') - n(0,\vk')) = \bar n (s,k) - n_0(k),
\end{equation}
where
\begin{equation}
	n_0(k) = \frac{\delta(k-\kcrit)}{k}.
\end{equation}
On the RHS we have
\[   
\begin{split}
	&\int_{0}^{\infty} \dd{q} q \int \dd{\vq'} \int \dd{\vk'}  \frac{\delta(k-k')}{k} \frac{\delta(q-q')}{q}  \frac{ \cov(\vk' - \vq')}{s^2 + T(k,q)^2} \\
	& \times \left ( \bar n (s,\vq') - \bar n (s,\vk') \right ),
\end{split}
\]
where we added the identity $\int_{0}^{\iy} \dd{q} q  \delta(q - q')/q = 1$. 
Calculating the integrals gives 
\[ \bar n (s, k) - n_0(k) = \int_{0}^{\infty} \dd{q} q \frac{2 \eps^2 \cov(k,q)}{s^2 + T(k,q)^2} \Gd \bar n (s,q,k),\]
where $\Gd \bar n (s,q,k) = \bar n (s,q) - \bar n (s,k)$. 

Sending $s \to 0$ gives 
\begin{empheq}[box=\ovalbox]{equation}
	\label{eq:first-limit}
	\bar  n(k) - n_0(k) =  \int_{0}^{\infty} \dd{q}q \frac{2 \eps^2 \cov(k,q) }{T(k,q)^2} \left ( \bar n(q) - \bar n(k) \right ),
\end{empheq}
where we have adapted the notation $\bar n (s=0,k) = \bar n(k)$. 

Changing into scaled energy variables $E = \f 1 2 k^2/ \eps$, $E' = \f 1 2 q^2/ \eps$ we can write $T(k,q) = \eps(E - E')$ and expand the function $\bar n (q) = \bar n (E') $ around $E$ giving 
\begin{equation*}
	\bar n(E') - \bar n (E) = (E' - E) \p_E \bar n (E) + \f 1 2 (E' - E)^2 \p^2_{E} \bar n (E) + \ldots
\end{equation*}
Dividing by $T^2 = \eps^2 (E -E')^2$ leaves us with terms of the form
\begin{equation}
	\frac{2}{(n+2)!} \int_{0}^{\iy} \dd{q} q\, \cov(k,q) (E'(q) - E)^{n} \p^{n+2}_E \bar n (E),
\end{equation}
with $n \geq -1$
on the RHS of Eq.~\eqref{eq:first-limit}. Writing the terms with $E$ and $E'$ in terms of $k$ and $q$ we have to evaluate
\[ \frac{2 \p_E^{n+2} \bar n}{2^{n}(n+2)!}\int_{0}^{\iy} \dd{q} q\, \cov(k,q) (q^2/\eps - k^2/\eps )^{n}.  \]
Because the covariance function $\cov$ is scaled by $\zeta$ we can write this in terms of scaled variables $Q = \zeta q$, $K = \zeta k$ giving
\begin{equation}
	\label{eq:nexpansion-tail}
	\frac{1}{(n+2)!} \frac{2 \p_E^{n+2} \bar n}{(2\eps \zeta^2)^{n}}\int_{0}^{\iy} \dd{Q} Q\, \cov_1(K,Q) (Q^2 - K^2 )^{n},
\end{equation}
where $\cov_\zeta (k,q) = \cov (k,q) $ is the covariance function with the scaling written down explicitly i.e. $\cov_1$ does not depend on $\zeta$ explicitly. 

We define $P(K,Q) = (K+Q)\cov_1(K,K+Q)$. We have to evaluate the magnitude of terms
\begin{equation}\label{eq:first-integral}
	\int_{-K}^{\iy} \dd{Q} P(K,Q) Q^n (2K + Q)^n,
\end{equation}
where, again,  $n \geq -1$. 
Notice that $\int_{-K}^{\iy} \dd{Q} P(K,Q) = 1$ since we had already integrated out the angle dependence. For this reason, this term will give a unity for $n=0$ contributing a term $\p_E^2 \bar n$. Since $\cov(\vk - \vq)$ is peaked at $\vk = \vq$, it is safe to assume that the angle averaged version $\cov_1(K,Q) = \int_{-\pi}^{\pi} \dd{\gt'} \cov_1(\vec K -\vec Q(\gt'))$ is also peaked around $Q = K$ and equivalently $P(K,Q)$ is peaked around $Q=0$. For this reason, the function $P(K,Q)$ gives a noticeable contribution only when $Q \ll K$. Therefore, we expand the term $(2K +Q)^n$. Since $P(K,Q)$ is nearly an even function with respect to $Q$, we maintain the zeroth order term if $n$ is even giving 
\[ (2K)^n \int_{-K}^{\iy} \dd{Q} P(K,Q) Q^n. \]
In case $n>0$ is odd, the integral is approximated by the first term 
\[ n(2K)^{n-1} \int_{-K}^{\iy} \dd{Q} P(K,Q) Q^{n+1}. \]
The contribution of odd $n$ terms is minimal compared to the even terms. This is seen by plugging in $K = \zeta k$ in the prefactor, which shows that the prefactor has a smaller power of $\zeta$ than the even $n$ terms.

Considering only the even $n$ terms, the truncation of the expansion of $\bar n$  depends on the magnitude of 
\begin{equation}\label{eq:braodening-n-powers}
	\frac{1}{(n+2)!} \frac{2k^n}{(\eps \zeta)^n} \int_{-K}^{\iy} \dd{Q} P(K,Q) Q^n,
\end{equation}
where we replaced $K = \zeta k$ in the prefactor. Since $P(K,Q)$ does not have explicit $\zeta$ dependence, the scaling is fully contained in the prefactor. 
It turns out that it suffices that $\eps \zeta \approx 1$: even for $n=2$ the integral gives the variance that is close to unity but the prefactor has the factorial $1/4! = 1/24$. For sufficiently peaked $P(K,Q)$ the moments should not grow faster than the factorial. However, it should be kept in mind that the large value of $\zeta k$ was needed earlier to ensure that the angle averaged $\cov_1$ is indeed peaked at $Q = K$. Since $n=0$ contributed a unity, the only remaining term is $n=-1$ corresponding to the singular integral
\begin{equation}\label{eq:singular-integral}
	4 \eps \zeta^2 \int_{-K}^{\iy} \dd{Q} \frac{P(K,Q)}{Q(2K + Q)},
\end{equation}
where we have retained the term $2K + Q$. 

In order to evaluate the magnitude of the integral Eq.~\eqref{eq:singular-integral} we have to go back to the definition of 
\begin{equation}
	\begin{split}
		&\frac{P(K,Q)}{K+Q} = \cov_1 (K,K+Q)  \\
		&= \int_{-\pi}^{\pi} \dd{\gt} \tilde \cov_1 (Q^2 + 2K(K+Q)(1-\cos \gt)),
	\end{split}
\end{equation}
where $\tilde \cov_1$ is the scaled covariance function with the argument written in terms of the squared modulus i.e. $\cov_1 (K,Q) = \tilde \cov_1 (|\vec K - \vec Q|^2)$ and $\gt $ is the angle between $\vec Q$ and $\vec K$. In order to correctly capture the scaling behavior we write this integral in terms of the circle line element $\diff s  = K \diff \gt$ having 
\[ \frac{P(K,Q)}{K+Q} 
= \frac{1}{K}\int_{-\pi K}^{\pi K} \dd{s} \tilde \cov_1 (Q^2 + 2K(K+Q)(1-\cos (s/K) )), \]
Since $K$ is assumed to be large, we expand the argument of the $\cos$ and discard the small terms $Q/K$. We have 
\[ P(K,Q) = \frac{K+Q}{K} \tilde P(K,Q), \]
where
\[ \tilde P(K,Q) \approx  \int_{-\pi K}^{\pi K} \dd{s} \tilde \cov_1 (Q^2 + s^2). \]
$\tilde P (K,Q)$ is clearly minimized when $Q = 0$ and is an even function of $Q$. Plugging this back in Eq.~\eqref{eq:singular-integral} gives
\[ 4 \eps \zeta^2 \int_{-K}^{\iy} \dd{Q} \frac{\tilde P(K,Q)}{Q} \frac{K+Q}{K(2K + Q) }. \]
We expand the latter rational expression in $Q$ giving
\[ 4 \eps \zeta^2 \int_{-K}^{\iy} \dd{Q} \frac{\tilde P(K,Q)}{Q} \frac{1}{2K} \left ( 1 + \frac{Q}{2K} \right ) + \mathcal{O}(\zeta^2 K^{-3}). \]
Since $\tilde P$ is even function of $Q$, the first term gives no contribution. $\tilde P$ still integrates approximately to unity resulting in a contribution 
\begin{equation}
	\frac{1}{ 2E} \p_E \bar n(E).
\end{equation}
The exact value of this term depends on the approximations we made but we will show that it does not matter for the solution to the lowest order in $\eps$ as the assumption $E \gg 1$ continues to apply. We have included the calculation for the approximate covariance function Eq.~\eqref{eq:covariance-continuous} in App.~\ref{sec:test-broadening} and it turns out that in that case this term scales as $1/\zeta^2$. 

We plug in our results in Eq.~\eqref{eq:first-limit} giving
\begin{equation}
	\label{eq:energy-tail}
	\bar n (\go) - \eps^2 \left ( \bar n'' (\go) + \frac{1}{2 \go} \bar n'(\go) \right ) = n_0(\go),
\end{equation}
where we have introduced the unscaled energy variable $\go = \eps E = \frac 1 2 k^2$. Here $n_0 (\go) = \delta (\go - \go_\crit ) $ with $\go_\crit = \f 1 2 \kcrit^2$ and $\bar n$ is subject to the normalization condition $\int_{0}^{\iy} \dd{\go} \bar n (\go) = 1$. Eq.~\eqref{eq:energy-tail} can be solved exactly by using solutions $\go^{1/4} I_{1/4} (\go/\eps)$ and $\go^{1/4} K_{1/4} (\go/\eps)$ (modified Bessel functions) to the homogeneous problem $n_0(\go) \to 0$. However, since we have already assumed that $\go \gg \eps$ we will solve the homogeneous problem by plugging in an ansatz
\begin{equation}
	\bar n (\go ) = e^{f(\go) /\eps },
\end{equation}
where
\begin{equation}
	f(\go) = \sum_{j=0}^{\iy} \eps^{j} f_j(\go)
\end{equation}
is expanded in $\eps$. This gives an equation 
\begin{equation}
	f'(\go)^2 + \eps f''(\go) + \eps f'(\go)/(2\go) = 1.
\end{equation}
To the zeroth order in $\eps$ this reads
\begin{equation}
	f_0'(\go)^2 = 1
\end{equation}
solved by $f_0(\go) = \pm \go + c$ ($c$ is a constant of integration). Then the homogeneous solutions are $\exp(\pm \go/\eps)$, which actually solve the differential equation 
\begin{equation}\label{eq:energy-tail-approximate}
	\bar n (\go) - \eps^2 \bar n ''(\go) = \delta(\go - \go_\crit ),
\end{equation}
when $\go \neq \go_\crit$. This is the reason for dropping the term with the single derivative -- it contributes to a correction in the exponential decay of the order $\mathcal O(1)$. The first order correction gives the homogeneous solution $\exp(\pm \go/ \eps)/\go^{1/4}$, which corresponds to the asymptotic large argument approximation of the earlier solutions expressed in terms of the modified Bessel functions. 

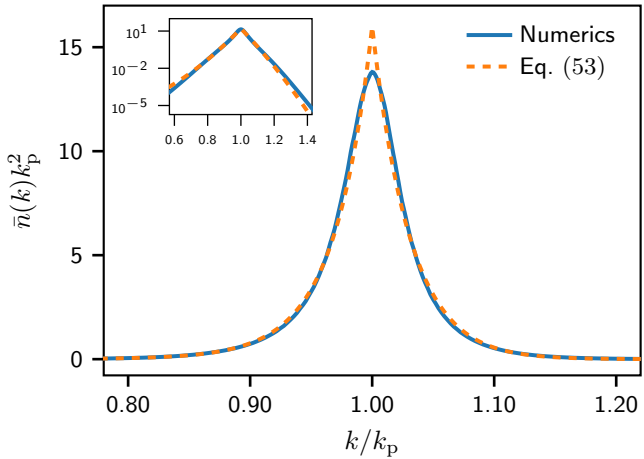
\begin{figure}
	\centering
\begin{tikzpicture}

\definecolor{darkgray176}{RGB}{176,176,176}
\definecolor{darkorange25512714}{RGB}{255,127,14}
\definecolor{lightgray204}{RGB}{204,204,204}
\definecolor{steelblue31119180}{RGB}{31,119,180}

\begin{axis}[
	width=\columnwidth,
	height=0.75\columnwidth,
	legend cell align={left},
	legend style={fill opacity=0.8, draw opacity=1, text opacity=1, draw=none, font={\small \sffamily}},
	x grid style={darkgray176},
	xlabel={$k/\kcrit$},
	xmin=0.78, xmax=1.22,
	ymin=-0.78, ymax=17,
	y grid style={darkgray176},
	ylabel={$\bar n(k) \kcrit^2$},
	x tick label style={/pgf/number format/.cd,
		fixed,
		fixed zerofill,
		precision=2,
		/tikz/.cd
	},
	every axis plot/.append style={line width=1.5pt}
]
\addplot [steelblue31119180] table {figures/data/radial_energy_data.txt};

\addlegendentry{Numerics}

\addplot [darkorange25512714, dashed] table {figures/data/radial_energy_fit.txt};
\addlegendentry{Eq.~\eqref{eq:decay-solution}}

\end{axis}

\begin{axis}[
	log basis y={10},
	xmin=0.5701, xmax=1.4279,
	ymin=1.864e-06, ymax=100,
	xshift=.1\columnwidth,
	yshift=.4\columnwidth,
	width=0.4\columnwidth,
	height=0.33\columnwidth,
	minor tick style={draw=none},
	major tick style={line width = .5pt},
	x tick label style = {font=\sansmath\sffamily\tiny,
		/pgf/number format/.cd,
		fixed,
		fixed zerofill,
		precision=1,
		/tikz/.cd,
	},
	y tick label style = {font=\sansmath\sffamily\tiny},
	axis line style = {line width=.5pt},
	ymode=log,
	every axis plot/.append style={line width=1.5pt}
	]
	\addplot [steelblue31119180] table {figures/data/radial_energy_data.txt};
	\addplot [dashed, darkorange25512714] table {figures/data/radial_energy_fit.txt};
\end{axis}

\end{tikzpicture}
	\caption{The profile of the long-time number density $\bar n(k)$ is plotted against numerical data. The inset shows the same data on a logarithmic scale revealing the asymmetric behavior around the peak. 
	The numerical data presented here is calculated for $\zeta = 12/\kcrit$ and $\eps = \kcrit^2/32 $ and averaged over 100 realizations. For details see App.~\ref{sec:numerical-methods}.
	}
	\label{fig:tail_vs_numerics}
\end{figure}

We solve the inhomogeneous problem Eq.~\eqref{eq:energy-tail-approximate} corresponding to the first term in the approximation discussed earlier. We take the Fourier transform giving
\begin{equation}
	 \hat n (\xi) + \eps^2 \xi^2  \hat n(\xi) = \int_\R \dd{\go} e^{-i \go \xi} \delta \left (\go - \ocrit \right )
	= e^{-i \xi \ocrit},
\end{equation}
which is solved by
\begin{equation}
	\bar n (\go) = \frac{1}{2\pi \eps} \int_{\R} \dd{\eta} \frac{e^{i \frac{\eta}{\eps} (\go - \ocrit ) }}{1 +  \eta^2}
	= \frac{1}{2\eps} \exp(-\frac{\left |\go - \ocrit \right |}{\eps}).
\end{equation}
We can add any solution to the homogeneous equation $\bar n (\go) - \eps^2 \bar n''(\go) = 0$ to fix the normalization. The final solution is
\begin{equation}
	\bar n (\go) = \frac{1}{2\eps} \left ( \exp(-\frac{\left |\go - \ocrit \right |}{\eps}) + \exp(-\frac{\go + \ocrit }{\eps}) \right ).
\end{equation}
In terms of $k$ this becomes
\begin{equation}
	\label{eq:decay-solution-real}
	\bar n (k) = \frac{1}{2\eps} \left ( \exp(-\frac{\left |k^2 - \kcrit^2 \right |}{2 \eps}) + \exp(-\frac{k^2 + \kcrit^2 }{2\eps}) \right ).
\end{equation}
However, since the integrated error in the normalization without the correction is of the order $\mathcal O (\exp(-\f 1 2 \kcrit^2 \eps^{-1}) )$, we have the approximate solution
\begin{empheq}[box=\ovalbox]{equation}
	\label{eq:decay-solution}
	\bar n (k ; \kcrit) = \frac{1}{2\eps} \exp(-\frac{\left |k^2 - \kcrit^2\right |}{2\eps}).
\end{empheq}
Fig.~\ref{fig:tail_vs_numerics} shows this result against direct numerical simulation of Eq.~\eqref{eq:schrodinger_dynamics} using the approximate covariance kernel defined in Eq.~\eqref{eq:covariance-approximation}.

Taking the logarithm $\log(\bar n(k))$ gives 
\begin{equation}
	\log(\frac{\bar n(k)}{\bar n (\kcrit)} ) 
	= - \frac{|k - \kcrit| \kcrit}{\epsilon } -  \operatorname{sgn}(k - \kcrit) \frac{(k - \kcrit)^2}{2 \eps}
\end{equation} 
around the peak $k = \kcrit$. 
This shows that the quadratic correction to the exponential decay rate switches sign at $k = \kcrit$ so while for $k > \kcrit $ the correction is superexponential, for $k< \kcrit$ the correction is subexponential. This observation is confirmed by the numerical results (Fig.~\ref{fig:tail_vs_numerics}). Another important observation is that the result does not depend on the correlation length $\zeta$. Instead, the broadening of the initial peak is fully determined by the potential energy scale $\eps$. We should also mention that the derivation above can be done with minor modifications in 3 dimensions giving the same leading term exponential decay. 

Finally, we note that we have in effect calculated the Green's function $\bar n (k;q)$,
which can be used to obtain a solution for more general initial conditions $n_0$ as 
\begin{equation}
	\label{eq:green-solution}
	\bar n (k) = \int_0^{\infty} \dd{q} q \bar n(k;q) n_0(q).
\end{equation}
If preserving the integrated density $\int_0^{\infty} \dd{k} k \bar n(k) $ is essential, one should use Eq.~\eqref{eq:decay-solution-real} for the Green's function. 
However, it should be kept in mind that the derivation was based on the assumption of separation of scales between the initial wave momentum $\kcrit$ and the inverse correlation length $\zeta^{-1}$. Therefore, Eq.~\eqref{eq:green-solution} is only applicable for initial conditions supported in regions of $k$ with $k \gg 1/\zeta$. 

\subsection{Angle dependence}\label{sec:angle_dependence}
In this Section we calculate the first order correction to the $s \to 0$ results discussed in the previous Section. The main goal is to give a long-time description of how the field $n(\vk)$  becomes isotropic. 

We start by calculating the Fourier coefficients of Eq.~\eqref{eq:laplace_transform} in the $\gt$ variable of the polar coordinates $\vk = k (\cos \gt, \sin \gt)$, $\vq = q (\cos \gt', \sin \gt')$. We write
\begin{equation}
	\bar n_m(s,k) = \int_{-\pi}^{\pi} \dd{\gt} e^{-i m \gt} \bar n (s,\vk(\gt)).
\end{equation}
On the LHS of Eq.~\eqref{eq:laplace_transform} we have $ \bar n_m (s,k) - n_{0,m}(k)$ and the RHS gives
\[ \int_{0}^{\infty} \dd{q} q \int_{-\pi}^{\pi} \dd{\gt'} \int_{-\pi}^{\pi} \dd {\gt} e^{-i m \gt} \frac{2 \epsilon^2 \cov (k,q,\gt-\gt')}{s^2 + T(k,q)^2} \Delta \bar n (s,\vq,\vk) . \]
Calculating the integrals and multiplying by $s$ on both sides gives
\begin{equation}
	\label{eq:laplace-fourier}
	\begin{split}
		s \bar n_m &(s,k) - s n_{0,m}(k) = 	
		2\pi \eps^2 \int_{0}^{\infty} \dd{q}q \, \delta_s\left  (T(k,q)\right ) \\ & 
		\times \left (  \cov_m(k,q)  \bar n_m(s,q) -  
		\cov_0(k,q)  \bar n_m(s,k) \right ),
	\end{split}
\end{equation}
where
\begin{equation}
	\delta_s \left (T(k,q)\right ) = \frac{s/\pi}{s^2 + T(k,q)^2},
\end{equation}
\begin{equation}
	\cov_m (k,q) = \int_{-\pi}^{\pi} \dd{\gt} e^{-im\gt} \cov (k,q,\gt),
\end{equation}
and
\begin{equation}
	n_{0,m} (k) = \int_{-\pi}^{\pi} \dd{\gt} e^{-im\gt} n_0(\vk(\gt)).
\end{equation}

Next we assume that $s$ is small and expand this equation in $s$ writing
\begin{equation}
	\label{eq:n-expansion}
	\bar n_m(s,k) =  \bar n_m^{(0)}(k) + s \bar n_m^{(1)}(k) + \mathcal O (s^2).
\end{equation}
In order to expand $\delta_s$, we write Eq.~\eqref{eq:laplace-fourier} in energy coordinates $\go = \f 1 2 k^2$, $\go' = \f 1 2 q^2$. We examine the functional
\begin{equation}
	\int_{0}^{\infty} \dd{\go'} \delta_s(\go-\go') f(\go',\go),
\end{equation}
where we have suppressed the $s$ dependence of $f$. 
The zeroth order term gives the Dirac delta $\delta_s (\go) \to \delta (\go)$,
while the first order term gives 
\begin{equation}
	\begin{split}
		\lim_{s \to 0} \frac{1}{\pi}\int_{0}^{\infty} \dd{\go'} &\left ( \p_s \frac{s}{s^2 + (\go-\go')^2} \right ) f(\go',\go) 
		\\&= \frac{1}{\pi} \int_{0}^{\infty} \dd{\go'} \frac{f(\go',\go)}{(\go-\go')^2}.
	\end{split}
\end{equation}
A necessary condition for this integral to be finite (as a Cauchy principal value) is that $\lim_{\go' \to \go} f(\go',\go) = 0$. To summarize, we have
\begin{equation}\label{eq:deltas_expansion}
	\delta_s(\go) = \delta(\go) + \frac{s}{\pi \go^2} + \mathcal{O}(s^2). 
\end{equation}

Plugging in the expansion Eq.~\eqref{eq:n-expansion} and the identity Eq.~\eqref{eq:deltas_expansion} above in Eq.~\eqref{eq:laplace-fourier} gives to zeroth order 
\[ \int \dd{q} q \delta \left ( T(k,q) \right ) 
\left  (  \cov_m(k,q) \bar n_m^{(0)} (q)  -  \cov_0(k,q) \bar n_m^{(0)} (k) \right ) = 0, \]
which evaluates to 
\begin{equation}
	\bar n_m^{(0)}(k) \left (  \cov_m(k) -  \cov_0(k) \right ) = 0,
\end{equation}
where we use the notation $\cov_m(k,k) \to \cov_m(k)$.
Since $ \cov_m(k) <  \cov_0(k)$ (see App.~\ref{sec:monotonicity}) when $m>0$, this is true only if $ \bar n_m^{(0)}(k) = 0$ when $m \neq 0$ i.e. the field $\bar n$ becomes isotropic as $s \to \iy$.

The first order equation reads 
\begin{equation}
	\begin{split}
		&\bar n_m^{(0)}(k) - n_{0,m}(k) = 2\pi \eps^2 \left (  \cov_m(k) -  \cov_0(k) \right )  \bar n_m^{(1)} (k) \\
		&+ \int \dd{q}  \frac{2 \eps^2 q}{T(k,q)^2} 
		\left (  \cov_m (k,q)  \bar n_m^{(0)}(q) -  \cov_0 (k,q)  \bar n_m^{(0)}(k) \right ).
	\end{split}
\end{equation}
Now if $m = 0$, this gives the equation for the $s \to 0$ limit discussed in Sec.~\ref{sec:momentum-broadening}, namely
\begin{equation*}
	\bar n_0^{(0)}(k) - n_{0,0}(k) = \int \dd{q} q \frac{2 \eps^2 \cov_0 (k,q)}{T(k,q)^2} 
	 \left (   \bar n_0^{(0)}(q) - \bar n_0^{(0)}(k) \right ).
\end{equation*}
If $m \neq 0$ the first order correction is
\begin{equation}\label{eq:first_order_correction}
	\bar n_m^{(1)} (k) = \frac{n_{0,m}(k)}{2\pi \eps^2 \left ( \cov_0(k) -  \cov_m(k)\right )}.
\end{equation}
Next, we calculate $\cov_m$ for the approximate covariance Eq.~\eqref{eq:covariance-continuous}:
\begin{equation}
	\begin{split}
		\cov_m(k) &=  \int_{-\pi}^{\pi} \dd{\gt} e^{-i m \gt} \cov(k,\gt)  \\
		&= \frac{ \zeta^2}{2\pi} \int_{-\pi}^{\pi} \dd{\gt} e^{-i m \gt}  e^{-\zeta^2 k^2 \left ( 1-\cos \gt \right )}.
	\end{split}
\end{equation}
Since $\zeta k \gg 1$, we can use Laplace's method expanding the argument of the $\cos$ giving
\begin{equation}
	\begin{split}
		\cov_m(k)
		&\approx \frac{\zeta^2}{2\pi} \int_{-\pi}^{\pi} \dd{\gt} e^{-i m \gt}  e^{-\frac{\zeta^2 k^2}{2} \gt^2 } \\
		&\approx \frac{ \zeta^2}{2\pi} \int_{-\infty}^{\infty} \dd{\gt} e^{-i m \gt}  e^{-\frac{\zeta^2 k^2}{2} \gt^2}.
	\end{split}
\end{equation}
Replacing the limits $(-\pi,\pi)$ with $(-\infty,\infty)$ is justified if $\zeta k \gg 1$. 
Calculating the Fourier transform results in 
\begin{equation}\label{eq:diffusion_covariance_function}
	\cov_m(k) \approx \frac{ \zeta  }{k \sqrt{2 \pi }} e^{-\frac{m^2}{2 \zeta ^2 k^2}}.
\end{equation}

Plugging Eq.~\eqref{eq:diffusion_covariance_function} in Eq.~\eqref{eq:first_order_correction} gives
\begin{equation}
	\bar n_m^{(1)} (k) = \frac{k n_{0,m}(k)}{\sqrt{2\pi} \zeta \eps^2 \left (1 - \exp(-\f 1 2 \frac{m^2}{\zeta^2 k^2})\right )}.
\end{equation}
For small $m$ we have 
\[  \bar n_m^{(1)} (k) = \frac{2 k^3 \zeta n_{0,m}(k)}{\sqrt{2\pi} \eps^2 m^2 } \]
while for large $m$ 
\[   \bar n_m^{(1)} (k) = \frac{k n_{0,m}(k)}{\sqrt{2\pi} \zeta \eps^2 }.\]

We find that for the slowest modes to relax we must have 
\begin{equation}
	s \ll \frac{\eps^2}{\zeta k^3} \sqrt{\frac{\pi}{2}}.
\end{equation}

\section{Boltzmann equation}\label{sec:boltzmann}
The linear Boltzmann equation on the $k$-space ring has been derived rigorously \cite{Spohn1977,Erdos2000} in the kinetic scaling limit $T = \eps^2 t$, $X = \eps^2 x$ as $\eps \to 0$ when the dimension $d \geq 2$. We shall rewrite the result here assuming the truncation of the cumulant hierarchy described earlier. 
We start with Eq.~\eqref{eq:laplace_transform} denoting the Laplace transform as $\nlaplace$:
\begin{equation}\label{eq:laplace_dynamics}
	s \nlaplace(s,\vk) - n_0(\vk) = \int \dd{\vq}  \frac{2  \eps^2 s \cov(\vk - \vq)}{s^2 + T(k, q)^2} \Gd \nlaplace(s,\vq,\vk).
\end{equation}
We define the rescaled field in variables $S = s/\eps^2$, $T = \eps^2 t$ writing $N_\eps (T,\vk) = n(t,\vk) $. The Laplace transform becomes $\nlaplace (s,\vk) = \tilde N_\eps(S,\vk)/\eps^2 $. Since $s \nlaplace = S \tilde N_\eps$, the rescaled equation reads
\begin{equation}\label{eq:laplace_dynamics_rescaled}
	S \tilde N_\eps (S,\vk) - N_0(\vk) = \int \dd{\vq}  \frac{2 \eps^2  S \cov(\vk - \vq)}{\eps^4 S^2 + T(k, q)^2} \Gd \tilde N_\eps (S,\vq,\vk).
\end{equation}
Letting $\eps \to 0$ gives the previously discussed limit
\begin{equation}
	\frac{\eps^2 S}{\eps^4 S^2 + T(k, q)^2} \to \pi \delta\left ( T(k,q) \right ).
\end{equation}
Calculating the inverse transformation gives 
\begin{empheq}[box=\ovalbox]{equation}
	\label{eq:diffusion-equation}
	\p_t n(t,\vk) = 2 \pi \eps^2 \int \dd{\vq} \delta\left (T(k,q) \right )  \cov(\vk -\vq)  \Gd n(t,\vq,\vk).
\end{empheq}
in terms of the original field corresponding to the lowest order expansion in $\eps$. 

Next we write Eq.~\eqref{eq:diffusion-equation} in polar coordinates $\vk = k (\cos \gt, \sin \gt)$, $\vq = k (\cos \gt', \sin \gt')$ on the ring $q = k$. 
Taking the Fourier transform $\int_{-\pi}^{\pi} \dd{\gt} e^{-im\gt}$ of Eq.~\eqref{eq:diffusion-equation} gives
\begin{equation}\label{eq:diffusion_coefficient}
	\p_t n_m(t,k) = 2\pi \eps^2 \left ( \cov_m(k) - \cov_0 (k) \right ) n_m (t,k),
\end{equation}
where we used the convolution theorem 
\begin{equation}
	\int_{-\pi}^{\pi} \dd{\gt} e^{-im \gt} \int_{-\pi}^{\pi} \dd{\gt'} f(\gt-\gt') g(\gt') = f_m g_m.
\end{equation}

\begin{figure}\label{fig:beginning_scaling}
	\centering
\begin{tikzpicture}

\definecolor{darkgray176}{RGB}{176,176,176}
\definecolor{darkorange25512714}{RGB}{255,127,14}
\definecolor{lightgray204}{RGB}{204,204,204}
\definecolor{steelblue31119180}{RGB}{31,119,180}

\begin{groupplot}[group style={group size=2 by 1,horizontal sep=6pt},
height=0.8\columnwidth,
width=\columnwidth]
\nextgroupplot[
xlabel={$10^2 \eps/\kcrit^2$},
ylabel={$10^2/(t_{\text{c}} \kcrit^2)$},
width=.6\columnwidth,
height=0.48\columnwidth,
legend cell align={left},
legend style={
  fill opacity=1,
  fill=none,
  draw opacity=1,
  draw=none,
  text opacity=1,
  at={(0.03,0.97)},
  anchor=north west,
  font={\scriptsize \sffamily},
},
tick align=outside,
tick pos=left,
x grid style={darkgray176},
xmin=2.375, xmax=5.125,
ymin=1, ymax=8,
xtick style={color=black},
y grid style={darkgray176},
ytick style={color=black}
]
\addplot [semithick, steelblue31119180,mark=*]
table {%
2.5000e+00 1.8306e+00
3.0000e+00 2.5973e+00
3.5000e+00 3.5665e+00
4.0000e+00 4.6295e+00
4.5000e+00 5.8089e+00
5.0000e+00 7.1857e+00
};
\addlegendentry{numerics}

\addplot [semithick, darkorange25512714]
table {figures/data/epsilon_fit.txt};
\addlegendentry{fit $\propto \eps^2$}

\nextgroupplot[
xlabel={$\zeta \kcrit$},
width=.6\columnwidth,
height=0.48\columnwidth,
legend cell align={left},
legend style={
	fill opacity=1,
	fill=none,
	draw opacity=1,
	draw=none,
	text opacity=1,
	at={(0.03,0.97)},
	anchor=north west,
	font={\scriptsize \sffamily},
},
tick align=outside,
tick pos=left,
x grid style={darkgray176},
xmin=5.75, xmax=10.25,
ymin=1, ymax=8,
xtick style={color=black},
y grid style={darkgray176},
yticklabel=\empty,
]
\addplot [semithick, steelblue31119180,mark=*]
table {%
6.0000e+00 2.5892
6.8000e+00 2.9459
7.6000e+00 3.2694
8.4000e+00 3.5882
9.2000e+00 3.9345
1.0000e+01 4.2395
};
\addlegendentry{numerics}
\addplot [semithick, darkorange25512714]
table {%
6.0000e+00 2.5700
6.8000e+00 2.9127
7.6000e+00 3.2554
8.4000e+00 3.5980
9.2000e+00 3.9407
1.0000e+01 4.2834
};
\addlegendentry{fit $\propto \zeta$}
\end{groupplot}

\end{tikzpicture}
	\caption{The initial peak decays as $\propto e^{-t/t_\text{c}} $, where $t_\text{c}$ is given by Eq.~\eqref{eq:decay_rate}. The numerical results show the correct scaling with $\eps$ and $\zeta$. In the first plot $\zeta = 12/\kcrit$ and fitting $1/t_{\text{c}} = C_\eps \eps^2$ gives $C_\eps = 0.96$ times the theoretical prediction. In the second plot $\eps = 1/24 \kcrit^2$ and $1/t_\text{c} = C_\zeta \zeta$; fitting gives $C_\zeta=0.98$ times the theoretical value. In both cases each point is calculated by averaging over 1000 simulations.}
\end{figure}
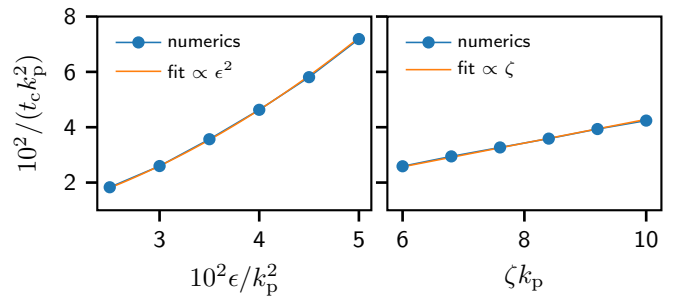

Plugging in the approximation Eq.~\eqref{eq:diffusion_covariance_function} in Eq.~\eqref{eq:diffusion_coefficient} gives
\begin{equation}\label{eq:diffusion_coefficient_final}
	\p_t n_m(t,k) = \frac{\sqrt{2 \pi } \zeta  \epsilon ^2 }{k} \left ( e^{-\frac{m^2}{2 \zeta ^2 k^2}} -1 \right ) n_m(t,k).
\end{equation}
The modes with small $m$ decay according to
\begin{equation}
	\p_t n_m(t,k) = -\frac{  \epsilon ^2  }{k^3 \zeta} \sqrt{\frac{\pi}{2}} m^2 n_m(t,k)
\end{equation}
equivalent to the diffusion equation
\begin{equation}\label{eq:diffusion-approximated}
	\partial_{t} n(t,k,\gt) = \frac{\eps^2}{k^3 \zeta} \sqrt{\frac{\pi}{2}} \p_\gt^2 n(t,k,\gt).
\end{equation}
The dissipation of large $m$ modes defines a collision time-scale
\begin{equation}\label{eq:decay_rate}
	\frac{1}{t_\text{c}} = 2\pi \eps^2 \cov_0 (k) \approx \sqrt{2 \pi} \zeta \eps^2/k,
\end{equation} 
results that we already saw in the end of Sec.~\ref{sec:angle_dependence}. Fig.~\ref{fig:beginning_scaling} shows the latter decay rate obtained from simulating Eq.~\eqref{eq:schrodinger_dynamics}. 

The slowest mode corresponding to $m=1$ defines another, diffusive, time scale
\begin{equation}\label{eq:diffusive-time-scale}
	\frac{1}{t_{\text{d}}} = 2\pi \eps^2 (\cov_0(k) - \cov_1(k)) \approx \sqrt{\frac{\pi}{2}} \frac{\eps^2}{\zeta k^3 }.
\end{equation}
The time $t_\text{d}$ is the time for a wave packet to forget its initial direction and is an important characteristic of the diffusive regime. Authors in Ref.~\cite{Kuhn2005,Kuhn2007} find the same relations for $t_\text{c}$ and $t_\text{d}$ expressed in terms of Fourier coefficients $\cov_m(k)$, albeit using notedly different methods. 

\section{Diffusion equation}\label{sec:diffusion}
A Boltzmann equation can be derived in a similar way to Eq.~\eqref{eq:diffusion-equation} for the Wigner quasiprobability distribution function (Wigner function) \cite{Erdos2008}
\begin{equation}
	W(\vx,\vk) = \frac{1}{(2\pi)^d} \int \dd{\xi} e^{i \xi \cdot \vx} \psi^*\left ( \vk - \frac{\xi}{2} \right ) \psi \left ( \vk + \frac{\xi}{2} \right )
\end{equation}
yielding 
\begin{equation}\label{eq:wigner-boltzmann}
	D_t F = L F,
\end{equation}
where
\begin{equation}
	\begin{split}
		L F(t,\vx,\vk) &= 2 \pi \eps^2 \int \dd{\vq}   \cov(\vk-\vq) \delta  ( T(k,q)  ) \\
		& \times \left ( F(t,\vx,\vq) -F(t,\vx,\vk) \right ).  
	\end{split}
\end{equation}
Here, $F(t,\vx,\vk) = \lan W(t,\vx,\vk) \ran$ and $D_t = \p_t + \vk \cdot \nabla_{\vx}$ is the advective derivative. $F$ can be seen as a quasiprobability distribution for finding the particle at position $\vx$ with momentum $\vk$. Importantly, it has the marginal $\int \dd{\vx} F(t,\vx,\vk) = n(t,\vk)$ and we see that the linear Boltzmann equation Eq.~\eqref{eq:diffusion-equation} can be obtained by integrating Eq.~\eqref{eq:wigner-boltzmann} over $\vx$. Let $\rho(t,\vx) = \int \dd{\vk} F(t,\vx,\vk)$ be the probability (or number density) of finding a particle at $\vx$. Authors of Ref.~\cite{Erdos2008} showed that $\rho$ obeys the diffusion equation
\begin{equation}
	\partial_t \rho(t,\vx) = D \nabla^2 \rho(t,\vx)
\end{equation}
in the scaling limit $t \to \eps^{2 + 2 \kappa}t $ , $x \to \eps^{2+\kappa}x$ with some small $\kappa > 0$ as $\eps \to 0$ for dimensions larger than 2. 

The diffusion constant $D$ can be obtained using a probabilistic interpretation of Eq.~\eqref{eq:wigner-boltzmann}, where 
\begin{equation}
	F(t,\vx,\vk) = \lan \delta (\vx - \vec X_t) \delta (\vk - \vec K_t) \ran_{\vec X, \vec K}
\end{equation}
is the probability density function of stochastic variables $\vec X_t$ and $\vec K_t$ obeying the dynamics 
\begin{equation}
	\p_t \vec X_t = \vec K_t
\end{equation}
and the time evolution of the probability density $n(t,\vk)$ of $\vec K_t$ is determined by Eq.~\eqref{eq:diffusion-equation} i.e.
\begin{equation}\label{eq:momentum-boltzmann}
	\p_t n = L n. 
\end{equation}
The dynamics of $\vec K$ can be seen as a Poisson process on the circle with radius $k$ and jump rate
\begin{equation}
	\gl(k) = 2\pi \eps^2 \bar \cov (k)
\end{equation}
and jump size probability distribution function
\begin{equation}
	\bar \cov (k)^{-1} \cov(\vk - \vq),
\end{equation}
where we have
\begin{equation}
	\bar \cov (k) = \int \dd{\vq} \delta (T(k,q)) \cov (\vk - \vq).
\end{equation}

As already pointed out in Ref.~\cite{Erdos2008}, the diffusion constant $D$ can be obtained using a Green-Kubo formula
\begin{equation}
	D = \frac{1}{d} \int_{0}^{\iy} \dd{t} \lan \vec K(t) \cdot \vec K(0) \ran,
\end{equation}
where $d$ is the dimension of the system and $\vec K$ is a Markov process defined by Eq.~\eqref{eq:momentum-boltzmann}, whose initial condition $\vec K(0)$ is sampled from the equilibrium distribution of the said Markov process. Here, the diffusion tensor is isotropic due to the isotropy of $L$, which is ultimately implied by the isotropic properties of $V$. 

We can calculate the diffusion constant in 2 dimensions using the polar coordinates $\vk = k (\cos \gt, \sin \gt)$. Since the equilibrium distribution of the momentum isotropization process is the uniform distribution on the circle with radius $k$, we get a particularly nice expression. In polar coordinates we have
\begin{equation}
	\begin{split}
		Ln(t,k,\gt) &= 2\pi \eps^2 \int_{-\pi}^{\pi} \dd{\gt'} \cov(k,\gt - \gt') \\
		& \times \left ( n(t,k,\gt') - n(t,k,\gt) \right )
	\end{split}
\end{equation}
and
\begin{equation}
	D = \f {k^2} 2 \int_{0}^{\iy} \dd{t} \frac{1}{2\pi} \int \dd{\gt_0} \int \dd{\gt} \cos(\gt - \gt_0) n(t,\gt),
\end{equation}
where $n(t,\gt)$ is the probability distribution with the initial condition $n_0(\gt) = \delta(\gt - \gt_0)$. This can be expressed as $n(t) = e^{tL} n_0$, allowing the calculation of the time integral. We notice that $L$ commutes with the shift operator $S(-\gt_0) f(\gt) = f(\gt - \gt_0)$ giving 
\begin{equation}
	D = \frac{k^2}{2} \int_{0}^{\iy} \dd{t} \int_{-\pi}^{\pi} \dd{\gt} \cos(\gt) e^{tL} \delta(\gt).
\end{equation}
Calculating the time integral results in
\begin{equation}\label{eq:diffusion-constant-2}
	D = -\frac{k^2}{2} \int_{-\pi}^{\pi} \dd{\gt} \cos(\gt) L^{-1} \delta (\gt).
\end{equation}
The operator $ L $ is diagonalized by the Fourier transform. We have the Fourier transform of $L n$
\begin{equation}
	L n_m(k) = \hat L_m(k) n_m(k),
\end{equation}
where 
\begin{equation}
	\hat L_m(k) = 2\pi \eps^2 ( \cov_m(k) - \cov_0(k)).
\end{equation}
Since $\cov(k,\gt)$ is an even function of $\gt$, Eq.~\eqref{eq:diffusion-constant-2} gives the Fourier transform
\begin{equation}
	D = -\frac{k^2}{2\hat L_1(k)} = \frac{k^2}{4\pi \eps^2 (\cov_0(k) - \cov_1(k))}.
\end{equation}
Comparing with Eq.~\eqref{eq:diffusive-time-scale} we write
\begin{equation}\label{eq:diffusion-constant-time}
	D = \frac{t_\text{d} k^2 }{2},
\end{equation}
which gives
\begin{equation}
	D \approx \frac{1}{\sqrt{2\pi}} \frac{\zeta k^5}{\eps^2}
\end{equation}
for the approximation Eq.~\eqref{eq:covariance-approximation}. Although derived using Green's function methods, the relationship Eq.~\eqref{eq:diffusion-constant-time} is also reported in Refs.~\cite{Kuhn2005,Kuhn2007}. 

\section{Conclusion and discussion}\label{sec:conclusion}
We have derived an expression for the momentum broadening  Eq.~\eqref{eq:decay-solution} in the long-time limit assuming that the potential energy scale $\eps$ is small compared to the kinetic energy of the initial wave. Furthermore, we calculated the general expressions for the diffusion time scale and the diffusion constant by using a Boltzmann equation applicable for small $\eps$. The diffusive time scale was also obtained from a direct expansion in the long-time limit. 

The results depend on the low-order truncation of the cumulant hierarchy. This truncation is expected to yield better results if $V$ is a Gaussian free field.
However, no such assumption is needed for the results presented here as long as $\eps$ is sufficiently small. 

It is known that the two-dimensional system is localized for any finite $\eps$ as the system size tends to infinity \cite{Abrahams1979}. Moreover, interference effects are also expected to hinder spatial diffusion due to wave-interference effects. Authors in Refs.~\cite{Kuhn2005,Kuhn2007} introduce a localization length 
\begin{equation}
	\xi_{\text{loc}}(k) = k t_\text{d} e^{\pi D} \approx \frac{\sqrt{\frac{\pi}{2}} k^4 \zeta }{\eps^2} \exp(\sqrt{\frac{\pi}{2}} \frac{\zeta k^5 }{\eps^2})
\end{equation}
that should be comparable to the system size $L$ in order to observe Anderson localization. We see immediately that in our case $\xi_{\text{loc}}$ is extremely large implying that even weak localization effects are negligible. 

The momentum broadening can be interpreted as a signature of the mixing mechanism that thermalizes the non-interacting gas. This interpretation is in the spirit of Eigenstate thermalization hypothesis \cite{Deutsch2018}. Introducing the random potential $V$ breaks the linear and angular momentum conservation of the free particles. However, since the gas is non-interacting we still expect to observe a kinetic energy close to that of the initial state. Calculating the expectation value
\begin{equation}
	\Tr (O \hat \rho) = \frac{1}{\mathcal N} \sum\nolimits_\vk n_\vk \bra{k} O \ket{k},
\end{equation}
for an operator $O$ we see that it is in fact a microcanonical trace operation, where $n_\vk$ picks up exactly the energies within $\eps$ from the initial kinetic energy $\f 1 2 \kcrit^2$. Here $\hat \rho$ is the density matrix and we have included $\mathcal N = \sum_\vk n_\vk$ in case $n$ is not normalized to unity. Another way to see this result is that the $t \to \infty$ limit gives exactly the thermalized state of one particle. The finite width $\eps$ gives now information about the mixing mechanism needed to thermalize the system. 

\section*{Acknowledgments}
The work has been supported by the Academy of Finland, via an Academy project (project No. 339228), the {\em Finnish Centre of Excellence in Randomness and Structures\/} (project Nos. 346306 and 364213) and {\em the Finnish Quantum Flagship} (project No. 358878).

\appendix


\section{Mathematical preliminaries}
\subsection{Cumulants} \label{sec:cumulants}
Let $I=(i_1, i_2, \ldots, i_n)$ be a list $n$ indices. We define the moment of a collection of stochastic variables as
\begin{equation}\label{eq:moment_definition}
	\langle X^{I} \rangle = \left \langle \prod_{j \in I} X_j \right \rangle,
\end{equation} 
where $\langle \cdot \rangle $ denotes the average over the joint probability measure $\mu(X_I) = \mu(X_{i_1},X_{i_2},\ldots,X_{i_n})$. We will use the shorthand $X_{k^*} = X_{k}^*$, where the last term means the complex conjugate of $X_k$.

Cumulants of a list of stochastic variables $X_I$ can be defined using the moments to cumulants formula
\begin{equation}\label{eq:moments2cumulants}
	\langle X^{I} \rangle = \sum_{\pi \in \mathcal P (I) } \prod_{A \in \pi} \kappa(X_{A}).
\end{equation}
This is the sum over all partitions $\pi$ of the index set $ I $, where the product is taken over different \emph{clusters} $A$ in the partition $\pi$. 

\paragraph{Example:} Let $I = (1,2,3)$. The partitions in this case would be $\{ (1,2,3)\}$, $\{ (1,2), (3) \}$, $\{ (1,3), (2) \}$, $\{ (2,3), (1) \}$, and $\{ (1),(2), (3) \}$. This means that the sum in Eq.~\eqref{eq:moments2cumulants} would have 5 terms altogether. 

The time evolution of cumulants can be written using the following formula:
\begin{equation}\label{eq:cumulant_time_evolution}
	\partial_{t} \kappa (X_I) = \sum_{j \in I} \langle \partial_t X_j : X^{I\setminus j} : \rangle,
\end{equation}
where $:X^{I \setminus j}:$ is the Wick polynomial of the stochastic variables with the index $j$ excluded. It suffices to know that Wick polynomials have the following property:
\begin{equation}
	\label{eq:expansion-wick}
	\langle X^{K} : X^{I} : \rangle = \sum_{\pi \in \mathcal P (I + K) } \prod_{J \in \pi} \chi(J \cap K \neq \emptyset ) \kappa(X_J),
\end{equation}
where $\chi$ evaluates to 1 when its argument is true and to 0 otherwise. Here we use a notation in which $I + K$ gives the concatenation of lists $I$ and $K$. The meaning of Eq.~\eqref{eq:expansion-wick} is that any partitions with clusters internal to $X^{I}$ in the cluster expansion of the expectation value $\langle X^{K} : X^{I} : \rangle$ are excluded. 

\paragraph{Example:}
For monomials $X^{K} = X_0$ we have 
\begin{equation}
	\langle X_0 :X^{I}: \rangle = \kappa (X_0,X_{i_1},X_{i_2},\ldots,X_{i_N}) = \kappa(X_0,X_I),
\end{equation}
whereas $X^K = X_1 X_2$ gives
\begin{equation}
	\begin{split}
		\lan X_1 X_2 : X^{I} : \ran &= \kappa(X_1,X_2,X_I)  \\
		&+ \sum_{A \subset I} \kappa(X_1, X_{A}) \kappa(X_2, X_{I \setminus A} ),
	\end{split}
\end{equation}
where the sum goes through all the subsets $A$ of $I$ including the empty set $\emptyset$.
We refer the reader to Ref.~\cite{Lukkarinen2016} for more details on Wick polynomials and their relation to cumulants.

\subsection{Homogeneous isotropic random fields} \label{sec:random_fields}
We will be looking at properties of random field $\psi(\vx)$, where $\vx \in \mathbb{T}(L_1,L_2,\ldots, L_d)$ i.e. a box with sizes $L_i$ with periodic boundaries. We write the Fourier coefficients by defining 
\begin{equation}
	\psi(\vx) = \sum_{\vk \in \Lambda_d} \psi_\vk \exp(i \vk \cdot \vx),
\end{equation}
where $\Lambda_d$ denotes the set of reciprocal lattice points
\begin{equation}
	\Lambda_d = \left\lbrace \left(n_1 \Delta k_1, \ldots, n_d \Delta k_d \right) : \left(n_1,  \ldots, n_{d}\right) \in \Z^{d}) \right\rbrace
\end{equation} 
and $\Delta k_i = 2\pi/L_i$. 

We say that the field is \emph{homogeneous} if
\begin{equation}\label{eq:homogeneous_field}
	\langle \psi(\vx) \psi(\vx + \vec{d}_2) \ldots \psi(\vx + \vec{d}_{n}) \rangle = \cov_n(\vec{d}_2,\ldots, \vec{d}_{n} ),
\end{equation}
i.e. the  $n$-point covariance function $\cov_n$ is independent of $\vx$. 
This means that the spatial correlations do not depend on the base point $\vx$ i.e. the field is statistically translation invariant. Writing this in terms of the Fourier coefficients gives
\begin{equation}
	\begin{split}
		&C_n (\vec{d}_2,\vec{d}_3,\ldots, \vec{d}_{n}) \\
		&= \sum_{\vk_j \in \Lambda_d} \langle \psi_{\vk_1} \ldots \psi_{\vk_n} \rangle e^{i (\vk_1 + \ldots + \vk_n) \cdot \vx } e^{i (\vk_2 \cdot \vec{d}_2 + \ldots + \vk_{n} \cdot \vec{d}_{n} ) }.
	\end{split}
\end{equation}
This is true for all $\vx $ and $\vec{d}_j$ if and only if the resonance condition
\begin{equation}\label{eq:resonance_real}
	\langle \psi_{\vk_1} \ldots \psi_{\vk_n} \rangle = 0, \; \text{if} \; \sum_{j=1}^{n} \vk_j  \neq 0
\end{equation}
holds. In case of complex fields we can let any of the fields in Eq.~\eqref{eq:homogeneous_field} be complex conjugated and the resulting resonance condition reads
\begin{equation}\label{eq:resonance_complex}
	\langle \psi_{\vk_1}^{(s_1)} \ldots \psi_{\vk_n}^{(s_n)} \rangle = 0, \; \text{if} \; \sum_{j=1}^{n} s_j \vk_j  \neq 0,
\end{equation}
where 
\begin{equation}
	\psi_{\vk_j}^{(s_j)} = \begin{cases}
		\psi_{\vk_j} & \text{if $s_j =1$};\\
		\psi_{\vk_j}^* & \text{if $s_j =-1$},
	\end{cases}
\end{equation}
i.e. the sign in the resonance condition is flipped for the fields that have been complex conjugated. 

Two very useful corollaries are  
\begin{equation}
	\langle \psi_\vk \rangle = 0,
\end{equation} 
when $\vk \neq 0$ and 
\begin{equation}
	\langle  \psi_{\vk}^* \psi_{\vq} \rangle = 0, 
\end{equation}
if $\vk \neq \vq$. The same resonance condition holds for cumulants. This can be seen by using the cumulants to moments formula 
\begin{equation}\label{eq:cumulants2moments}
	\kappa (  \psi_I ) = \sum_{\pi \in \mathcal{P}(I)} (|\pi|-1)! (-1)^{|\pi|-1} \prod_{A \in \pi} \langle  \psi^{A} \rangle,
\end{equation}
where the sum goes over all partitions $\mathcal P(I)$ of $I$ and $|\pi|$ is the number of clusters in the partition $\pi$. If $\sum_{\vk \in I} \vk \neq 0 $, it follows that for each term in the sum in Eq.~\eqref{eq:cumulants2moments}, at least one $A \in \pi$ has the property $\sum_{\vk \in A} \vk \neq 0 $.

The homogeneous field $\psi$ is isotropic if the $n$-point correlation function $C_n (\vec{d}_2,\vec{d}_3,\ldots, \vec{d}_{n})$ is independent of simultaneous rotations of the vectors $\vec d_j$. In particular it means that the two-point correlation function $C_2(\vec d)$ only depends on the modulus $|\vec d|$. 

\section{Cumulant expansion of the time evolution}\label{app:cumulant_expansion}
%
The time evolution fields $\phi_\vk$ follow Eq.~\eqref{eq:schrodinger_kspace}. However, in this section we will include the parameter $\eps$ in the potential $V$ as $\eps V \to V$. We will use the multi-index notation defined in Sec.~\ref{sec:cumulants}, where multi-indices are comprised of vectors in $\vk$-space. 
We allow vectors of the form $\vk^*$, which is a shorthand defined by
\begin{equation}
	 \phi_{\vk^*} :=  \phi_\vk ^*.
\end{equation}
We use Eq.~\eqref{eq:cumulant_time_evolution} to write down the time evolution for a generic cumulant with $I,J\ne \emptyset$ as 
\begin{equation}
	\begin{split}
		&\partial_t \kappa ( \phi_{I},  V_J ) =
		\sum_{\vk \in I} \langle \partial_t  \phi_\vk :  V^J  \phi^{I \setminus \vk} : \rangle \\
		&=
		\sum_{\vk \in I} \sum\nolimits_{\vq}  K_{kq}^{(s_\vk)}(t) \langle  \phi_{\vq}^{(s_\vk)}  V_{\vk - \vq}^{(s_\vk)} :  V^J  \phi^{I \setminus \vk} : \rangle,
	\end{split}
	\label{eq:quantum-cumulant1}
\end{equation}
where, as before, $f^{(s_\vk)}$ is complex conjugated only if $\phi_\vk$ is complex conjugated.
Using Eq.~\eqref{eq:expansion-wick} gives
\begin{equation}
	\begin{split}
		&\partial_t \kappa ( \phi_{I},  V_J ) \\&=
		\sum_{\vk \in I} \sum\nolimits_{\vq} K^{(s_\vk)}_{kq}(t) \left[ \kappa( \phi_\vq^{(s_\vk)},  V_{\vk - \vq}^{(s_\vk)}, \phi_{I \setminus \vk}, V_J)\right. \\
		& + \sum_{\substack{K_1 \subset I\setminus \vk, \\ K_2 \subset J }} \left.\kappa( \phi_\vq^{(s_\vk)}, \phi_{K_1},  V_{K_2}) \kappa( V_{\vk - \vq}^{(s_\vk)}, \phi_{K_1^c},  V_{K_2^c}) \right],
	\end{split}
	\label{eq:quantum-cumulant2}
\end{equation}
where $K_i^{c}$ is the complement of the list $K_i$. 
Next, we integrate Eq.~\eqref{eq:quantum-cumulant2} from $0$ to $t$. Note that there are no correlations of $( \phi_I, V_J)$ at time $t=0$. This gives 
\begin{equation}
	\label{eq:quantum-cumulant-duhamel}
	\begin{split} 
		&\kappa ( \phi_{I},  V_J ) \\&= 
		\sum_{\vk \in I} \sum\nolimits_{\vq} \int_{0}^{t} \diff t_1 K_{kq}^{(s_\vk)}(t_1) \left[ \kappa( \phi_\vq^{(s_\vk)},  V_{\vk - \vq}^{(s_\vk)}, \phi_{I \setminus \vk}, V_J)\right. \\
		& + \sum_{\substack{K_1 \subset I\setminus \vk, \\ K_2 \subset J }} \left.\kappa( \phi_\vq^{(s_\vk)}, \phi_{K_1},  V_{K_2}) \kappa( V_{\vk - \vq}^{(s_\vk)}, \phi_{K_1^c},  V_{K_2^c}) \right].
	\end{split}
\end{equation}
We can use this equation to hierarchically expand any further cumulants. High order cumulants are generally small because
with every step of the expansion, a term $V$ is added introducing another small factor $\epsilon$. The resonance condition will pick up exactly one term in the downward expansion. The upward cumulant hierarchy can be truncated due to the small coupling parameter $\epsilon$. For details we refer the reader to Ref.~\cite{Erdos2008} and references therein.

In the following we will repeat the calculation to obtain the time evolution of $n_\vk = \gk(\phi_\vk,\phi_\vk^*)$. We should note that the truncation of the cumulant hierarchy works better for random potentials that have (near) Gaussian statistics. The reason for this is that if $V$ is Gaussian random field, $\kappa(V_I) = 0$ if the list $I$ contains more than 2 elements. Consequently Gaussian random fields are fully characterized by the first two cumulants $\kappa(V_\vk)$ and $\gk(V_\vk,V_\vq)$. However, for a sufficiently small $\eps$ the truncation of the hierarchy should work for random fields $V$ that have reasonably small fluctuations \cite{Erdos2008}. 

In order to calculate $\p_t n_\vk$ we had to expand the cumulant $\gk(V_{\vk -\vq},\phi_\vq,\phi_\vk^*)$ (see Sec.~\ref{sec:averaged_dynamics}). Using Eq.~\eqref{eq:quantum-cumulant-duhamel} we have now $I = (\vq,\vk^*)$ and $J = (\vk -\vq)$. Discarding the upward expansion (higher order cumulants), leaves the downward expansion, where both lists $K_1$ and $K_2$ are empty or have a single term. We see that $K_1$ cannot be empty because of the added phase degree of freedom. Also, since $\gk(V_\vk) = 0$, $K_2$ has to be empty. Summing over the terms in $I$ gives
\begin{equation}
	\begin{split}
		&\gk(V_{\vk-\vq}, \phi_\vq, \phi_\vk^*) \\
		&= \int_{0}^{t} \dd{t'} \sum\nolimits_\vp \left [ K_{qp} \gk(\phi_\vp, \phi_\vk^*) \right . \gk(V_{\vk-\vq},V_{\vq-\vp}) \\
		& \left . + K_{kp}^* \gk(\phi_\vp^*,\phi_\vq) \gk(V_{\vk - \vq}, V_{\vk-\vp}^*) \right ] + \text{h.o.},
	\end{split}
\end{equation}
where h.o. denotes the higher order cumulants that we will discard. We use the resonance condition Eq.~\eqref{eq:resonance_real} for the potential terms giving
\begin{equation}
	\gk(V_{\vk-\vq}, \phi_\vq, \phi_\vk^*) \approx \cov_{\vk - \vq} \int_{0}^{t} \dd{t'} K_{kq}^*(t') (n_\vq - n_\vk),
\end{equation}
which is equivalent to Eq.~\eqref{eq:triad_evolution} (we defined $V$ in this section s.t. it includes $\eps$). 

\section{Momentum broadening for the Gaussian covariance function}
In this section we repeat the calculations in Sec.~\ref{sec:momentum-broadening} for the 2-dimensional Gaussian covariance function given by Eq.~\eqref{eq:covariance-continuous}. We start by calculating the rescaled covariance function $\cov_1 (K,Q)$ discussed in Sec.~\ref{sec:momentum-broadening}. 

We can write Eq.~\eqref{eq:polar-covariance} for the rescaled covariance function as
\begin{equation}
	\cov_1 (K,Q) = \frac{e^{-K^2/2 - Q^2/2}}{2\pi}\int_{-\pi}^{\pi} \dd{\gt} e^{K Q \cos \gt},
\end{equation}
where $\gt$ is the angle difference between $\vec K$ and $\vec Q$. Calculating the integral gives 
\begin{equation}\label{eq:covariance-1-1}
	\cov_1 (K,Q) = e^{-\f 1 2 (K^2 + Q^2)} I_0(KQ),
\end{equation}
where $I$ is the modified Bessel function of the first kind. Since $Q \approx K$ and $K = \zeta k$ is assumed to be large, we can use an asymptotic approximation \cite{abramowitz+stegun}
\begin{equation}
	I_0 (K Q) = \frac{1}{ \sqrt{2\pi  KQ}} e^{K Q }\left( 1 + \mathcal O \left(\f 1 {K Q}\right) \right).
\end{equation} 
Plugging in the lowest order expansion in Eq.~\eqref{eq:covariance-1-1} gives 
\begin{equation}
	C_1 (K,Q) = \frac{1}{ \sqrt{2\pi  KQ}} e^{-\f 1 2 (Q-K)^2}.
\end{equation}

In Sec.~\ref{sec:momentum-broadening} we defined the function
\begin{equation}
	\begin{split}
		P(K,Q) &= (K+Q)\cov_1(K,K+Q) \\
		&= \sqrt{\frac{K+Q}{K}} \frac{1}{\sqrt{2\pi}}e^{-\f 1 2 Q^2}.
	\end{split}
\end{equation} 
We will first evaluate the singular term in the series expansion \eqref{eq:nexpansion-tail} given by 
\begin{equation}
	4 \eps \zeta^2 \int_{-\iy}^{\iy} \dd{Q} \frac{P(K,Q)}{Q(2K+Q)},
\end{equation}
where we have taken the lower limit to $-\iy$ ($K \gg 1$). 
We expand the function 
appearing in front of the normalized Gaussian giving
\begin{equation}
	\begin{split}
		\frac{1}{Q(2K+Q)}\sqrt{\frac{K+Q}{K}}
		&= \frac{1}{2 KQ} -\frac{Q}{16 K^3} \\&+ \frac{Q^2}{16 K^4} + \mathcal O \left ( \frac{Q^3}{K^5} \right ).
	\end{split}
\end{equation}
Notice that the constant $Q$ term is exactly zero in this expansion. Since the Gaussian $e^{-\f 1 2 Q^2}$ is an even function, only even powers of $Q$ give a contribution. The lowest order contribution is 
\begin{equation}
	4 \eps \zeta^2 \int_{-\iy}^{\iy} \dd{Q} \frac{Q^2}{16 K^4} \frac{1}{\sqrt{2\pi}} e^{-\f 1 2 Q^2} = \frac{\eps}{4 \zeta^2 k^4}
\end{equation}
showing that it scales with $1/(\zeta k)^2$ and is therefore small. 

Next we will show for the Gaussian covariance function that the expansion terms $n \geq 1$ are indeed small. To the lowest order we have $P(K,Q) = \exp(-\f 1 2 Q^2)/\sqrt{2\pi}$. We evaluate Eq.~\eqref{eq:braodening-n-powers} giving
\begin{equation}
	\frac{(n-1)!!}{(n+2)!} \frac{2 k^n}{(\eps \zeta)^n} = \frac{1}{2^{n/2} \frac{n}{2}! (n+1)(n+2)} \frac{2 k^n}{(\eps \zeta)^n}
\end{equation}
for even $n$.

\section{$\cov_0$ is the largest Fourier coefficient}
\label{sec:monotonicity}
The function $\cov(\vk - \vq) = \cov(k,q,\gt-\gt')$ is a function of the squared modulus $|\vk -\vq|^2 = k^2 + q^2 + 2 \cos(\gt - \gt') $. 
The Fourier coefficient $\cov_m(k)$ is defined as 
\begin{equation}
	\begin{split}
		\cov_m (k) &= \int_{-\pi}^{\pi} \dd{\gt} e^{-im\gt} \cov(k,\gt) \\
		&= 2 \int_{0}^{\pi} \dd{\gt} \cos(m \gt) \cov(k,\gt).
	\end{split}
\end{equation}
The latter equality follows from the fact that $\cov(k,\gt)$ is an even function of $\gt$. Since $\cov(k,\gt) \geq 0$, Hölder's inequality gives
\begin{equation}
	\begin{split}
		\cov_m(k) &= 2 \int_0^{\pi} \dd{\gt} \cos(m \gt) \cov(k,\gt) \\
		&<  2 \int_0^{\pi}\dd{\gt} |\cov(k,\gt)| \sup_{x \in (0,\pi)} |\cos (m x)| \\
		& =  \cov_0(k,q)
	\end{split}
\end{equation}
for all $m > 0$. The strict inequality follows from the fact that $\cos(m \gt) $ is not identically equal to 1 when $m>0$. 

\section{Numerical methods}\label{sec:numerical-methods}
We solved the time and space discretized version Eq.~\eqref{eq:schrodinger_dynamics} by using an operator splitting method introduced in \cite{Strang1968}. We will use space and time units $x_0 = 2\pi/L$, $t_0 = m x_0^2 / \hbar $, where $L \times L$ is the size of the periodic domain. We will use the same symbols for the coordinates throughout this Appendix with the understanding that they describe the system in these new units. In the new units the size of the periodic domain is $2\pi \times 2\pi$. 

We used $N \times N$ spatial points to describe the system and set the initial wave packet $\vk_\crit = (\kcrit,0)$. The spatial grid is then $\vx =  (j_x \Gd x,j_y \Gd x)$, where $\Gd x = 2\pi/N$ and $j_x,j_y \in \{0,\ldots, d-1 \}$. The Fourier domain is $(k_x,k_y)$, where $k_x$ and $k_y$ are integers between $\lfloor -\frac{N-1}{2} \rfloor$ and $\lfloor \frac{N-1}{2} \rfloor $. 
The time step was set to $\tau = 1/\kcrit^2$, which is one half of the kinetic energy of the initial wave packet. We write the Hamiltonian of the system $H = T + \eps V$, where $T$ is the kinetic energy operator with the $k$-space representation $T(k) = \f 1 2 k^2$. The time step is calculated as 
\begin{equation}\label{eq:splitting-method}
	\psi_{t+\tau} = e^{-i\tau K/2} e^{-i \tau \eps V} e^{-i\tau K/2} \psi_t.
\end{equation}
The discrete Fourier transforms
\begin{subequations}\label{eq:discrete-fourier-transform}
	\begin{align}
		\hat f(\vk) &= (\mathcal F f)(\vk) = \sum\nolimits_{\vx} f(\vx) e^{-i \vk \cdot \vx}; \\
		f(\vx) &= (\mathcal F^{-1} f)(\vk) = \frac{1}{N^2} \sum\nolimits_\vk \hat f(\vk) e^{i \vk \cdot \vx},
	\end{align}
\end{subequations}
are calculated using the Fast Fourier Transform.
The discretized version of Eq.~\eqref{eq:splitting-method} becomes 
\begin{equation}
	\hat \psi_{t + \tau}(\vk) = e^{-i\tau K/2} \mathcal F e^{-i \tau \eps V} \mathcal F^{-1}e^{-i\tau K/2} \hat \psi_t(\vk).
\end{equation}

The random field $V(\vx)$ is generated in $k$-space with the help of the covariance function Eq.~\eqref{eq:covariance-approximation}.
We then define the potential as
\begin{equation}\label{eq:collision-time}
	V(\vx) = d^2 \operatorname{Re} \mathcal{F}^{-1} \sqrt{\hat \cov(\vk) }( X(\vk) + i Y(\vk) ),
\end{equation}
where $X,Y \in \R^{N \times N}$ are discrete fields, whose elements are independently sampled from a normal distribution with zero mean and unit variance. The factor $N^2$ is needed for proper normalization of $\lan V(x)^2 \ran$ when using the transforms defined by Eq.~\eqref{eq:discrete-fourier-transform} (remember that $\sum_\vk \hat \cov (\vk) = 1$). 

\subsection{Momentum broadening}\label{sec:test-broadening}
For the simulations shown in Fig.~\ref{fig:tail_vs_numerics} we used $d = 384$ and set $\kcrit = 120$. The simulations ran for $2 \cdot 10^6$ time steps ($2 \cdot 10^6 \tau$). An output is generated every $10^3$ steps. We calculate $n(t,\vk) = |\psi(t,\vk)|^2$ and average over the second half of the simulation when the field has relaxed sufficiently. The averaging is done over time and 100 separate simulations. 

The angle averaged data shown in Fig.~\ref{fig:tail_vs_numerics} is calculated as 
\begin{equation}
	n(x_j) = \sum\nolimits_\vk g(x_j,\vk) n(\vk),
\end{equation}
where
\begin{equation}
	g(x,\vk) = C_g(x) \exp(- \f 1 2 (k(\vk) - x)^2/x_0^2)/k(\vk).
\end{equation}
The constant $C_g(x)$ is normalization factor s.t. $\sum_\vk g(x,\vk) = 1$ and the parameter $x_0$ defines the width of the averaging. For our calculation we used $x_0 = 0.5$, which corresponds to a half of the discretization size of $\vk$. For the evaluation points $x_j$ we use 300 points with uneven discretization having more points around the peak $x \approx \kcrit$. The data $n(x_j)$ in Fig.~\ref{fig:tail_vs_numerics} is normalized s.t. $\sum_j \diff x_j n(x_j) = 1$. Here $\diff x_j$ is the difference $x_{j+1}-x_{j}$. 

We should note that we see backscattering and forward scattering peaks in our simulations but the averaged value of $n$ does not change more that $9 \%$ as a function of the angle. 

\subsection{Collision time simulations}
For these simulations we used $d=256$ and $\kcrit = 64$. For each point in Fig.~\ref{fig:beginning_scaling} we simulate the system for 100 time steps and record $n(t,\vk_\crit)$ averaging over 1000 realizations. We make a linear fit to $(t,\log n(t,\vk_\crit))$  obtaining the decay rate $1/t_\text{c}$. Then a constant is fitted against $1/(t_\text{c} \eps^2)$ in the first case ($C_\eps$; $\zeta \kcrit = 12$), and $1/(t_\text{c} \zeta)$ in the second case ($C_\zeta$; $\eps/\kcrit^2 = 1/24 $). 

\bibliography{bibliography}

@article{Labeyrie2012,
url = {https://dx.doi.org/10.1209/0295-5075/100/66001},
year = {2012},
month = {dec},
publisher = {EDP Sciences, IOP Publishing and Società Italiana di Fisica},
volume = {100},
number = {6},
pages = {66001},
author = {Labeyrie, Guillaume and Karpiuk, Tomasz and Schaff, Jean-François and Grémaud, Benoît and Miniatura, Christian and Delande, Dominique},
title = {Enhanced backscattering of a dilute Bose-Einstein condensate},
journal = {Europhys. Lett.}
}

@article{Deutsch2018,
	doi = {10.1088/1361-6633/aac9f1},
	url = {https://doi.org/10.1088/1361-6633/aac9f1},
	year = {2018},
	month = {jul},
	publisher = {IOP Publishing},
	volume = {81},
	number = {8},
	pages = {082001},
	author = {Deutsch, Joshua M},
	title = {Eigenstate thermalization hypothesis},
	journal={Rep. Prog. Phys.}
}

@article{Kuhn2005,
  url = {https://doi.org/10.1103/PhysRevLett.95.250403},
  title={Localization of matter waves in two-dimensional disordered optical potentials},
  author={Kuhn, RC and Miniatura, Christian and Delande, Dominique and Sigwarth, Olivier and Mueller, Cord A},
  journal={Phys. Rev. Lett.},
  volume={95},
  number={25},
  pages={250403},
  year={2005},
  publisher={APS}
}

@article{Petrov2004low,
	title={Low-dimensional trapped gases},
	author={Dmitry S. Petrov and Dimitri M. Gangardt and Georgy V. Shlyapnikov and Georgy V. Shlyapnikov},
	journal={J. Phys. IV},
	year={2004},
	volume={116},
	pages={5-44},
	url={https://api.semanticscholar.org/CorpusID:13879640}
}

@book{Goodman2007speckle,
  title={Speckle phenomena in optics: theory and applications},
  author={Goodman, Joseph W},
  year={2007},
  publisher={Roberts and Company Publishers},
  doi = {https://doi.org/10.1117/3.2548484}
}

@article{Chin2010feshbach,
	title = {Feshbach resonances in ultracold gases},
	author = {Chin, Cheng and Grimm, Rudolf and Julienne, Paul and Tiesinga, Eite},
	journal = {Rev. Mod. Phys.},
	volume = {82},
	issue = {2},
	pages = {1225--1286},
	numpages = {0},
	year = {2010},
	month = {Apr},
	publisher = {American Physical Society},
	doi = {10.1103/RevModPhys.82.1225},
	url = {https://link.aps.org/doi/10.1103/RevModPhys.82.1225}
}

@article{Kondov2011three,
	author = {S. S. Kondov  and W. R. McGehee  and J. J. Zirbel  and B. DeMarco },
	title = {Three-Dimensional Anderson Localization of Ultracold Matter},
	journal = {Science},
	volume = {334},
	number = {6052},
	pages = {66-68},
	year = {2011},
	doi = {10.1126/science.1209019},
	URL = {https://www.science.org/doi/abs/10.1126/science.1209019}}

@article{Karpiuk2012,
	title = {Coherent Forward Scattering Peak Induced by Anderson Localization},
	author = {Karpiuk, T. and Cherroret, N. and Lee, K. L. and Gr\'emaud, B. and M\"uller, C. A. and Miniatura, C.},
	journal = {Phys. Rev. Lett.},
	volume = {109},
	issue = {19},
	pages = {190601},
	numpages = {5},
	year = {2012},
	month = {Nov},
	publisher = {American Physical Society},
	doi = {10.1103/PhysRevLett.109.190601},
	url = {https://link.aps.org/doi/10.1103/PhysRevLett.109.190601}
}

@article{Jendrzejewski2012,
	title = {Coherent Backscattering of Ultracold Atoms},
	author = {Jendrzejewski, F. and M\"uller, K. and Richard, J. and Date, A. and Plisson, T. and Bouyer, P. and Aspect, A. and Josse, V.},
	journal = {Phys. Rev. Lett.},
	volume = {109},
	issue = {19},
	pages = {195302},
	numpages = {5},
	year = {2012},
	month = {Nov},
	publisher = {American Physical Society},
	doi = {10.1103/PhysRevLett.109.195302},
	url = {https://link.aps.org/doi/10.1103/PhysRevLett.109.195302}
}

@inbook{Gorkov1996,
	author = {L. P. Gor'kov and A. I. Larkin and D. E. Khmel'nitskiĭ},
	title = {Particle conductivity in a two-dimensional random potential},
	booktitle = {30 Years of the Landau Institute — Selected Papers},
	chapter = {},
	pages = {157-161},
	doi = {10.1142/9789814317344_0022},
	publisher={World Scientific},
	year = {1996}
}

@article{Akkermans1985,
  url = {10.1051/jphyslet:0198500460220104500},
  title={Weak localization of waves},
  author={Akkermans, E and Maynard, R},
  journal={J. Physique Lett.},
  volume={46},
  number={22},
  pages={1045--1053},
  year={1985},
  publisher={Les Editions de Physique}
}

@article{Abrahams1979,
  title = {Scaling Theory of Localization: Absence of Quantum Diffusion in Two Dimensions},
  author = {Abrahams, E. and Anderson, P. W. and Licciardello, D. C. and Ramakrishnan, T. V.},
  journal = {Phys. Rev. Lett.},
  volume = {42},
  issue = {10},
  pages = {673--676},
  numpages = {0},
  year = {1979},
  month = {Mar},
  publisher = {American Physical Society},
  url = {https://link.aps.org/doi/10.1103/PhysRevLett.42.673}
}

@book{Baranovski2006,
  title={Charge transport in disordered solids with applications in electronics},
  author={Baranovski, Sergei},
  year={2006},
  publisher={John Wiley \& Sons},
  doi = {10.1002/0470095067}
}

@article{Vynck2023,
  title = {Light in correlated disordered media},
  author = {Vynck, Kevin and Pierrat, Romain and Carminati, R\'emi and Froufe-P\'erez, Luis S. and Scheffold, Frank and Sapienza, Riccardo and Vignolini, Silvia and S\'aenz, Juan Jos\'e},
  journal = {Rev. Mod. Phys.},
  volume = {95},
  issue = {4},
  pages = {045003},
  numpages = {63},
  year = {2023},
  month = {Nov},
  publisher = {American Physical Society},
  url = {https://link.aps.org/doi/10.1103/RevModPhys.95.045003}
}

@article{Anderson1958,
  title = {Absence of Diffusion in Certain Random Lattices},
  author = {Anderson, P. W.},
  journal = {Phys. Rev.},
  volume = {109},
  issue = {5},
  pages = {1492--1505},
  numpages = {0},
  year = {1958},
  month = {Mar},
  publisher = {American Physical Society},
  url = {https://link.aps.org/doi/10.1103/PhysRev.109.1492}
}

@article{Kuhn2007,
url = {https://dx.doi.org/10.1088/1367-2630/9/6/161},
year = {2007},
month = {jun},
publisher = {},
volume = {9},
number = {6},
pages = {161},
author = {Kuhn, R C and Sigwarth, O and Miniatura, C and Delande, D and Müller, C A},
title = {Coherent matter wave transport in speckle potentials},
journal = {New J. Phys.}
}

@article{Strang1968,
    author = {Strang, Gilbert},
    title = {On the Construction and Comparison of Difference Schemes},
    journal = {SIAM J. Num. Anal.},
    volume = {5},
    number = {3},
    pages = {506-517},
    year = {1968},
    URL = {https://doi.org/10.1137/0705041},
}

@article{Lukkarinen2016,
    author = {Lukkarinen, Jani and Marcozzi, Matteo},
    title = "{Wick polynomials and time-evolution of cumulants}",
    journal = {J. Math. Phys.},
    volume = {57},
    number = {8},
    pages = {083301},
    year = {2016},
    month = {08},
    issn = {0022-2488},
    url = {https://doi.org/10.1063/1.4960556}
}

@article{Plisson2013,
	author = {Plisson, T. and Bourdel, T. and M{\"u}ller, C. A.},
	doi = {10.1140/epjst/e2013-01756-8},
	id = {Plisson2013},
	isbn = {1951-6401},
	journal = {Eur. Phys. J. Spec. Top.},
	number = {1},
	pages = {79--84},
	title = {Momentum isotropisation in random potentials},
	url = {https://doi.org/10.1140/epjst/e2013-01756-8},
	volume = {217},
	year = {2013}
}

@article{Spohn1977,
	author = {Spohn, Herbert},
	date = {1977/12/01},
	id = {Spohn1977},
	isbn = {1572-9613},
	journal = {J. Stat. Phys.},
	number = {6},
	pages = {385--412},
	title = {Derivation of the transport equation for electrons moving through random impurities},
	url = {https://doi.org/10.1007/BF01014347},
	volume = {17},
	year = {1977}
}

@article{Erdos2008,
	author = {L{\'a}szl{\'o} Erdős and Manfred Salmhofer and Horng-Tzer Yau},
	title = {{Quantum diffusion of the random Schrödinger evolution in the scaling limit}},
	volume = {200},
	journal = {Acta Math.},
	number = {2},
	publisher = {Institut Mittag-Leffler},
	pages = {211 -- 277},
	year = {2008},
	URL = {https://doi.org/10.1007/s11511-008-0027-2}
}

@article{Erdos2000,
	author = {Erdős, László and Yau, Horng-Tzer},
	title = {Linear Boltzmann equation as the weak coupling limit of a random Schrödinger equation},
	journal = {Commun. Pure Appl. Math.},
	volume = {53},
	number = {6},
	pages = {667-735},
	doi = {{https://doi.org/10.1002/(SICI)1097-0312(200006)53:6<667::AID-CPA1>3.0.CO;2-5}},
	url = {https://onlinelibrary.wiley.com/doi/abs/10.1002/%28SICI%291097-0312%28200006%2953%3A6%3C667%3A%3AAID-CPA1%3E3.0.CO%3B2-5},
	year = {2000}
}

@book{abramowitz+stegun,
	address = {New York},
	author = {Abramowitz, Milton and Stegun, Irene A.},
	publisher = {Dover},
	title = {Handbook of Mathematical Functions with Formulas, Graphs, and Mathematical Tables},
	year = 1964
}

\end{document}